\documentclass[prd,aps,twocolumn,floats,floatfix,nofootinbib]{revtex4-1}

\usepackage[utf8]{inputenc}
\usepackage{amssymb,bm}
\usepackage{graphicx}
\usepackage{dcolumn}
\usepackage{bm}
\usepackage{graphics}
\usepackage{slashed}
\usepackage{enumerate}
\usepackage{amsmath}
\usepackage{url}
\usepackage[usenames,dvipsnames,svgnames,table]{xcolor}
\usepackage{color}
\usepackage{xcolor}
\usepackage{ulem}

\usepackage{changes}
\definechangesauthor[color=red]{FC}

\definecolor{color1}{RGB}{191, 0, 255}

\begin{document}

\title{Delayed jet launching in binary neutron star mergers with realistic initial magnetic fields}

\author{
	Ricard Aguilera-Miret$^{1}$,
	Carlos Palenzuela$^{2,3}$,
	Federico Carrasco$^{4}$,
	Stephan Rosswog$^{1,5}$
	Daniele Vigan\`o$^{6,7}$,
}

\affiliation{${^1}$University of Hamburg, Hamburger Sternwarte, Gojenbergsweg 112, 21029, Hamburg, Germany}
\affiliation{${^2}$Departament  de  F\'{\i}sica,  Universitat  de  les  Illes  Balears,  Palma  de  Mallorca, E-07122,  Spain}
\affiliation{$^3$Institute of Applied Computing \& Community Code (IAC3),  Universitat  de  les  Illes  Balears,  Palma  de  Mallorca, E-07122,  Spain}
\affiliation{$^4$Instituto de F\'isica Enrique Gaviola, CONICET-UNC, 5000 C\'ordoba, Argentina}
\affiliation{$^{5}$The Oskar Klein Centre, Department of Astronomy, AlbaNova, Stockholm University, SE-106 91 Stockholm, Sweden}
\affiliation{$^6$Institut d'Estudis Espacials de Catalunya (IEEC), 08034 Barcelona, Spain}
\affiliation{$^{7}$Institute of Space Sciences (IEEC-CSIC), E-08193 Barcelona, Spain}

\begin{abstract}
We analyze a long-lived hyper-massive neutron star merger remnant  (post-merger lifetime $>250$ ms) that has been obtained via large eddy simulations with a gradient subgrid-scale model. We find a clear helicoidal magnetic field structure that is governed by the toroidal component of the magnetic field. Although no jet emerges during the simulation time, we observe at late times a significant increase of the poloidal component of the magnetic field at all scales. 
We also compare with the results of several binary neutron star simulations with moderate resolution of $120$~m, that are evolved up to $50$~ms after the merger, which differ in terms of the initial topology and strength of the magnetic field. We find that the best choice is an isotropic small-scale magnetic field distribution that mimics the turbulent state that generically develops during the merger. This initial configuration reaches a closer agreement with our high-resolution  simulation results than the purely dipolar large-scale fields that are commonly employed in these type of simulations.
This provides a recipe to perform such simulations avoiding the computationally expensive grids required to faithfully capture the amplification of the magnetic field by Kelvin-Helmholtz instabilities.
\end{abstract}

\maketitle

%%%%%%%%%%%%%%%%%%%%%%
\section{Introduction}
%%%%%%%%%%%%%%%%%%%%%%

The detection of a binary neutron star (BNS) merger through a gravitational wave (GW) signal and an electromagnetic (EM) counterpart from the GW170817 event~\cite{LVC-BNS,LVC-MMA} (with Ligo and Virgo interferometers) marked the beginning of the multimessenger era. The short Gamma Ray Burst (sGRB) detected $\sim 1.7$ ms after the merger is compatible with the prediction of a launching jet~\cite{LVC-GRB,goldstein2017,savchenko2017,Troja2017,Margutti2017,Hallinan2017,Alexander2017,Mooley2018a,Lazzati2018,Lyman2018,Alexander2018,Mooley2018b,Ghirlanda2019}.
Moreover, the event confirmed earlier theoretical studies \cite{lattimer74,eichler89,rosswog99,freiburghaus99} that concluded that neutron star mergers must be major sources of r-process elements and the observed EM emission was compatible with the expected signatures of a  kilonova (e.g.,~\cite{Arcavi2017,Coulter2017,Pian2017,Smartt2017,Kasen2017,Metzger2019LRR} and references therein).

Currently, a realistic modelling of these phenomena relies mostly on numerical simulations solving the general relativistic magneto-hydrodynamical (MHD) equations for this specific scenario. The question of which processes drive the jet launching and what is the engine powering it are still unsolved problems, difficult to address even with the highest reachable resolutions and the most advanced numerical techniques~\cite{kiuchi15,palenzuela22,2024NatAs...8..298K,aguilera23}. Arguably, the most accepted hypothesis is the BH-disk scenario, in which the BNS merger produces either a hypermassive or a supramassive neutron star remnant that, after a short time, collapses to a black hole surrounded by an accretion disk (for a detailed discussion, see for instance the recent reviews~\cite{ciolfi2020key,Kiuchireview}). The magnetic field is amplified during and after the merger both in the neutron star's remnant and in the disk, where large-scale magnetic structures build-up through several mechanisms. The magnetic field flux is advected to the black hole horizon via the accretion process, accumulating and creating the necessary condition for the formation of a magnetically dominated jet~\cite{ruiz16}. Several studies suggested another possibility, commonly known as the magnetar hypothesis: an ultra-strongly
magnetized neutron star, instead of a black hole, drives the relativistic jet. Some numerical simulations possibly support this scenario ~\cite{ciolfi2020collimated,2024NatAs...8..298K}, but they consider unrealistically strong, purely poloidal large-scale magnetic fields as initial conditions in each neutron star, which might strongly bias the post-merger magnetic configuration (in particular, artificially enhancing large-scale structures), and therefore the conditions for jet formation.

Here we aim to shed some light on these questions by performing high-resolution Large-Eddy simulations (LES) of the long-lived hyper-massive neutron star (HMNS) produced during the merger of a relatively low-mass binary. Specifically, we extend a previous simulation~\cite{aguilera23} from $t=100$~ms up to $250$~ms, allowing us to study the development of the magnetic field in longer timescales. Despite reaching 
such (relatively) late times, we have not observed any jet emerging from the HMNS remnant. Our findings suggest that, when realistic magnetic strengths are considered and the remnant does not collapse to a black hole, large-scale magnetic field structures requires long timescales to form (i.e, of the order of few hundred milliseconds), and they are different from the often imposed pure dipolar field, since they arise from a slow inverse cascade that tend to organize the highly turbulent magnetic configuration produced after the collision.

In previous studies~\cite{palenzuela22,aguilera22} it was found that the average magnetic field strength is generically amplified from $B \lesssim 10^{12}$ to $10^{16}$~G in the bulk region of the remnant during the turbulent regime induced by the Kelvin-Helmholtz instability (KHI). Even with the improvement on the accuracy of the solution provided by LES, which effectively partially access sub-grid scales, the minimum resolution required in those simulations to achieve asymptotic convergence of the average magnetic field strength (i.e., it does not change increasing the numerical resolution) in the exponential magnetic field amplification phase was $\Delta x =60$~m (in the absence of LES, much smaller spacing is needed). Such high resolutions imply very expensive simulations, with a cost of $\mathcal{O}(10)$ million CPU-hours only to evolve the system for few tens of milliseconds after the merger. Facing this problem, many studies compensate the lack of resolution during the KHI induced amplification phase by starting with a strong poloidal large scale magnetic field before the merger.

In the present study, we perform several simulations, varying the strength and topology of the initial magnetic field, and compare them with our convergent high resolution results. As it was suggested in \cite{aguilera23}, the dynamics observed in high-resolution simulations starting with realistic magnetic fields can only be partially reproduced with under-resolved simulations if the seeded magnetic field is strong (average $B \sim 10^{15}$~G), but with an isotropic configuration with a distribution of energy over all scales, rather than the simple, often-used purely poloidal large-scale one.

The paper is organized as follows. Our numerical methodology,  including the evolution equations, numerical schemes, initial conditions, grid setup and analysis quantities, is briefly summarized in Section~\S\ref{sec:setup}. The numerical results of the long HMNS remnant evolution are presented and analyzed in Section~\S\ref{sec:long}. 
The impact of the initial strength and topology of the magnetic field in simulations under-resolving the MHD instabilities is studied in Section~\S\ref{sec:impact}.
Finally, our conclusions are presented in Section~\S\ref{sec:conclusions}.

%%%%%%%%%%%%%%%%%%%%%%%%%%%%%%%%%%%%%%%%%%%
\section{Numerical setup} \label{sec:setup}
%%%%%%%%%%%%%%%%%%%%%%%%%%%%%%%%%%%%%%%%%%%

%The concept and mathematical foundations behind LES with a gradient SGS approach have been extensively explained in our previous works (and references therein) in the context of classical~\cite{vigano19b} and relativistic MHD~\cite{carrasco19,vigano20,aguilera2020,palenzuela22,aguilera22}, to which we refer for details and further references. 
In the present work, we extend the simulation of the HMNS remnant presented in Ref.~\cite{aguilera23} by solving the full GRMHD equations combined with LES techniques. We also compare this simulation with lower resolution ones without LES. Initial data, evolution equations, numerical schemes and setup are almost identical to the ones employed in~\cite{aguilera23}, which we now summarize briefly for completeness.

%%%%%%%%%%%%%%%%%%%%%%%%%%%%%%%%%%%%%%%%%%%%%%%%%%%%%%%%%%%%%
\subsection{Evolution equations and numerical methods} \label{sec:equations}
%%%%%%%%%%%%%%%%%%%%%%%%%%%%%%%%%%%%%%%%%%%%%%%%%%%%%%%%%%%%%

The spacetime geometry is described by the Einstein equations, which can be written as an hyperbolic evolution system using the covariant conformal Z4 formulation~\cite{alic12,bezares17}. 
%A summary of the final set of evolution equations for the spacetime fields, together with the gauge conditions setting the choice of coordinates, can be found in~\cite{palenzuela18}.
The magnetized perfect fluid follows the GRMHD equations (see, e.g., \cite{shibatabook,palenzuela15}), a set of evolution equations for the conserved variables $ \left\lbrace D, U, S^i, B^i \right\rbrace$, which involve the mass, energy and momentum densities, together with the magnetic field. These conserved fields are functions of the primitive fields $ \left\lbrace \rho, \epsilon, v^{i}, B^i \right\rbrace$, namely the rest-mass density, the specific internal energy, the velocity vector and again the magnetic field. The recovery of the primitive from the conserved fields requires first a closure relation for the pressure $p$ as a function of the other thermodynamic fields (i.e., commonly known as the equation of state, EoS) and then solving a set of non-linear algebraic equations involving the Lorentz factor $W = (1-v^2)^{-1/2}$. We consider a hybrid EoS during the evolution, with a cold part modeled using a tabulated version of the piecewise polytrope fit to the APR4 zero-temperature EoS~\cite{read09,Endrizzi2016}. Thermal effects are modeled for simplicity by an additional ideal gas contribution $p_{\rm th} = (\Gamma_{\rm th} - 1) \rho \epsilon$, with adiabatic index $\Gamma_{\rm th}= 1.8$. The conversion from the evolved or conserved fields to the primitive or physical ones is performed by using the robust procedure introduced in~\cite{kastaun20}.
The full set of evolution equations, extended to the framework of LES and including all the gradient sub-grid-scale (SGS) terms, can be found in~\cite{vigano20,aguilera2020}. As in the first part of the simulation~\cite{aguilera23}, we include only the SGS term appearing on the induction equation with the pre-coefficient ${\cal C_M}= 8$, which has been shown to reproduce the magnetic field amplification more accurately (i.e., comparing with very high-resolution simulations) for our numerical schemes~\cite{vigano20,aguilera2020}.
We apply the SGS terms in the regions where the density is higher than $2 \times 10^{11}~\rm{g~cm^{-3}}$ in order to avoid possible spurious effects near the stellar surface and in the atmosphere (i.e., a
relatively tenuous, low-density fluid, which is necessary to prevent the failure of the HRSC schemes usually employed to solve the MHD equation).

% We remind the reader that these SGS terms, by construction, vanish at the continuous limit.  

%%%%%%%%%%%%%%%%%%%%%%%%%%%%%%
%\subsection{Numerical methods}
%%%%%%%%%%%%%%%%%%%%%%%%%%%%%%

As in our previous works, and in particular~\cite{aguilera2020,palenzuela22,aguilera22,aguilera23}, we use the code {\sc MHDuet}, generated by the platform {\sc Simflowny} \cite{arbona13,arbona18} and based on the {\sc SAMRAI} infrastructure \cite{hornung02,gunney16}, which provides the parallelization and the adaptive mesh refinement. Summarizing, it uses fourth-order-accurate operators for the spatial derivatives in the SGS terms and in the Einstein equations (the latter are supplemented with sixth-order Kreiss-Oliger dissipation); a high-resolution shock-capturing (HRSC) method for the fluid, based on the  Lax-Friedrich flux splitting formula \cite{shu98} and the fifth-order reconstruction method MP5 \cite{suresh97}; a fourth-order Runge-Kutta scheme with sufficiently small time step $\Delta t \leq 0.4~\Delta x$ (where $\Delta x$ is the grid spacing); and an efficient and accurate treatment of the refinement boundaries when sub-cycling in time~\cite{McCorquodale:2011,Mongwane:2015}. A complete assessment of the implemented numerical methods can be found in \cite{palenzuela18,vigano19}.

%%%%%%%%%%%%%%%%%%%%%%%%%%%%%%%
\subsection{Initial conditions and grid setup}
%%%%%%%%%%%%%%%%%%%%%%%%%%%%%%%

The initial data is created with the {\sc Lorene} package~\cite{lorene}, using the APR4 zero-temperature EoS described above. We consider an equal-mass BNS in quasi-circular orbit, with an irrotational configuration, a separation of $45$ km and an angular frequency of $1775\ \rm{rad~s^{-1}}$. The chirp mass $M_{chirp}=1.186~M_{\odot}$ is the one inferred in the refined analysis of GW170817 \cite{LVC-170817properties}, implying a total mass $M=2.724~M_{\odot}$ for the equal mass case. With this choice of EoS and masses, the remnant does not collapse to a black hole within the timespan of the long simulation, which is $250$ ms after the merger. 

The summary of the simulations performed to study the impact of the initial magnetic field is briefly described in Table~\ref{tab:models}. We refer to the different cases as {\tt BXXyyy}, where ${\tt BXX}$ corresponds to the average strength of the initial magnetic field and ${\tt yyy}$ represent the configuration, detailed as follows. In most of our simulations, marked as {\tt pol}, each star initially has a purely poloidal large-scale magnetic field that is confined to its interior, calculated from a vector potential toroidal component
\begin{equation}
	A_ {\phi} = A_0\,R^2 {\rm max}(p - p_{cut},0)
\end{equation}
where $p_{cut}$ is a hundred times the pressure of the atmosphere, and $R$ is the distance to the axis perpendicular to the orbital plane passing through the center of each star. We consider different values of the constant $A_0$, such that the averaged magnetic field within the star varies approximately between $10^{11}-10^{16}$~G. 
Alternatively, we use a very different, small-scale pseudo-isotropic magnetic field, marked as {\tt iso}, still confined to each stellar interior, by calculating the vector potential components as the product of harmonics with odd numbers 
\begin{eqnarray}
	A_ {x} &=& A_1 \sin(k_x x_c) \sin(k_y y_c) \cos(k_z z_c) \\
	A_ {y} &=& A_1 \sin(k_y x_c) \sin(k_z y_c) \cos(k_x z_c)
	\\
	A_ {z} &=& A_1 \sin(k_z x_c) \sin(k_x y_c) \cos(k_y z_c)
\end{eqnarray}
where $(x_c,y_c,z_c)$ are the coordinates with respect to the center of each star and we have defined the shortcut $A_1 =A_0 {\rm max}(p - p_{cut},0)$. We considered $(k_x,k_y,k_z)=(5,7,9)$ and a value of the constant $A_0$ such that the averaged quasi-isotropic magnetic field is $\sim 10^{15}$~G. 

These new simulations here presented are without the SGS (i.e., they are not LES) and use a lower resolution $\Delta x =120$~m. Therefore, they are not accurate enough to substantially resolve the KHI. We compare them with our previously mentioned reference, long LES simulation, which has a poloidal topology with an averaged strength $\sim 10^{11}$~G, several orders of magnitude lower than the large initial fields used in other simulations (e.g.,~\cite{kiuchi15,ruiz16,kiuchi18,ciolfi2019,ciolfi2020collimated,ruiz2020}) and not too far from the upper range of the oldest known neutron stars~\cite{bahramian23}. As it was shown in~\cite{aguilera22}, the initial topology is quickly forgotten after merging, at least as long as the initial values are not too large ($B\lesssim 10^{14}$ G) and the scheme and resolution are able to capture the turbulent amplification mechanisms.

By considering different strengths and topology of the magnetic field, we can confirm these previous results. More importantly, we provide a recipe to mimic the high-resolution results with moderate resources, namely by combining strong initial magnetic field strength with the quasi-isotropic topology.

\begin{table}[h]
	\begin{tabular}{ |c|c|c|c|c| }
		\hline
		Case
		& topology
		& $B_0$ [G]
		& Refinement levels
		& $\Delta x_{min}$ [m]
		\\ \hline
		{\tt MRLES} & poloidal & $10^{11}$ & 7 FMR + 1 AMR & 60--120 \\
		{\tt B16pol} & poloidal & $10^{16}$ & 7 FMR & 120 \\
		{\tt B15pol} & poloidal & $10^{15}$ & 7 FMR & 120 \\
		{\tt B15iso} & isotropic &$10^{15}$ & 7 FMR & 120 \\
		{\tt B14pol} & poloidal & $10^{14}$ & 7 FMR & 120 \\
		\hline
	\end{tabular}
	\caption{\textit{Parameters of the simulations:} Different simulations, with the initial topology and strength of the magnetic field, and the mesh refinement setup with the finest grid spacing $\Delta x_{min}$, indicating the maximum resolution.}
	\label{tab:models}
\end{table}

The binaries considered in this work are evolved in a cubic domain of linear size $\left[-1228,1228\right]$~km. The inspiral is fully covered by seven Fixed Mesh Refinement (FMR) levels. Each consists of a cube with twice the resolution of the next larger one, achieving a maximum resolution $\Delta x_{min} = 120$~m. This is the main setup for the medium LES presented in Section~\ref{sec:impact}.
For the long, high-resolution simulation presented in Section~\ref{sec:long}, we use one additional Adaptive Mesh Refinement (AMR) level for the first $t=100$~ms after the merger, covering the regions where the density is above $5 \times 10^{12}~\rm{g~cm^{-3}}$ and providing a uniform resolution throughout the shear layer. With this grid structure, we achieved a maximum resolution of $\Delta x_{min} = 60$~m covering at least the most dense region of the remnant for the first $t=100$~ms after the merger.  At that point, we assumed that such high-resolution is not required anymore to capture the relevant scales in the remnant's bulk region. Therefore, we remove the highest resolution level, evolving from $t=100$~ms to $t=250$~ms with a maximum resolution $\Delta x_{min} = 120$~m.
We focus the analysis in this paper on this new time interval of the simulation.

%%%%%%%%%%%%%%%%%%%%%%%%%%%%%%%%
\subsection{Analysis quantities}
%%%%%%%%%%%%%%%%%%%%%%%%%%%%%%%%

\begin{figure*}[ht!]
	\centering
	\includegraphics[width=0.35\linewidth, trim={0 4.5cm 0 0}, clip]{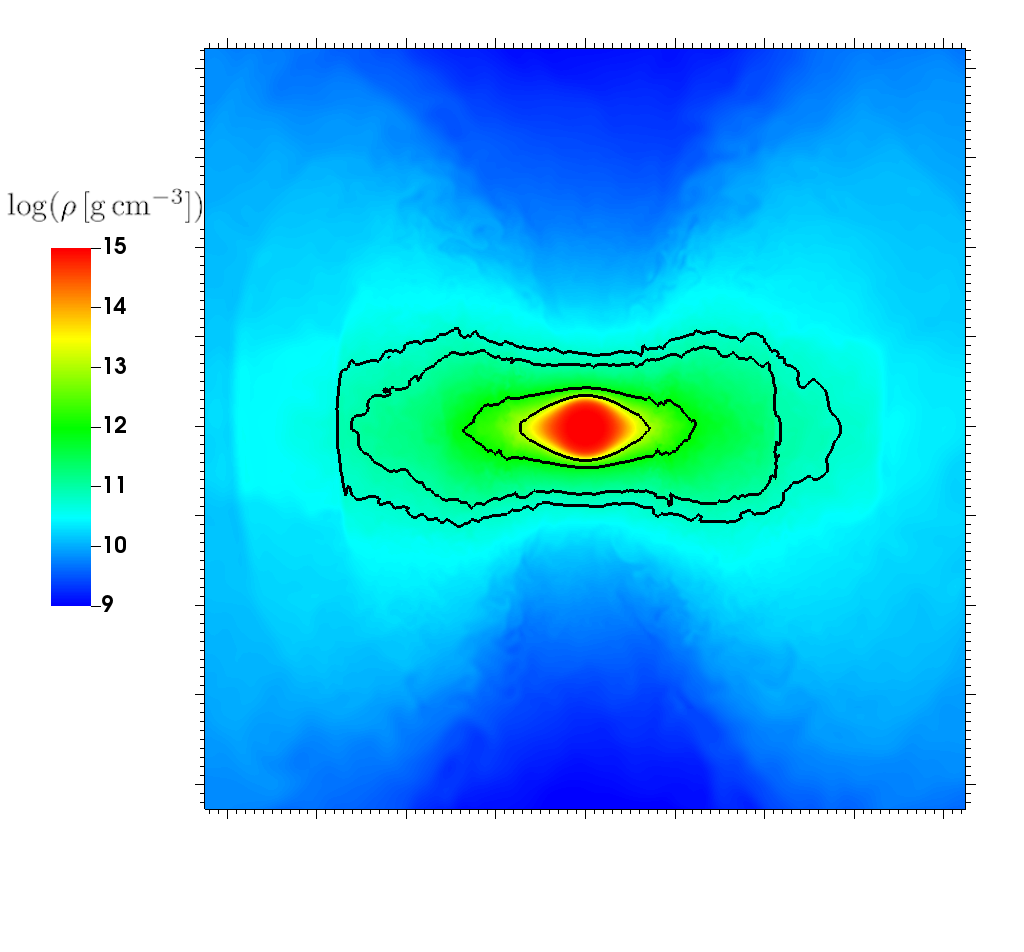}
	\includegraphics[width=0.3075\linewidth, trim={4.5cm 4.5cm 0 0}, clip]{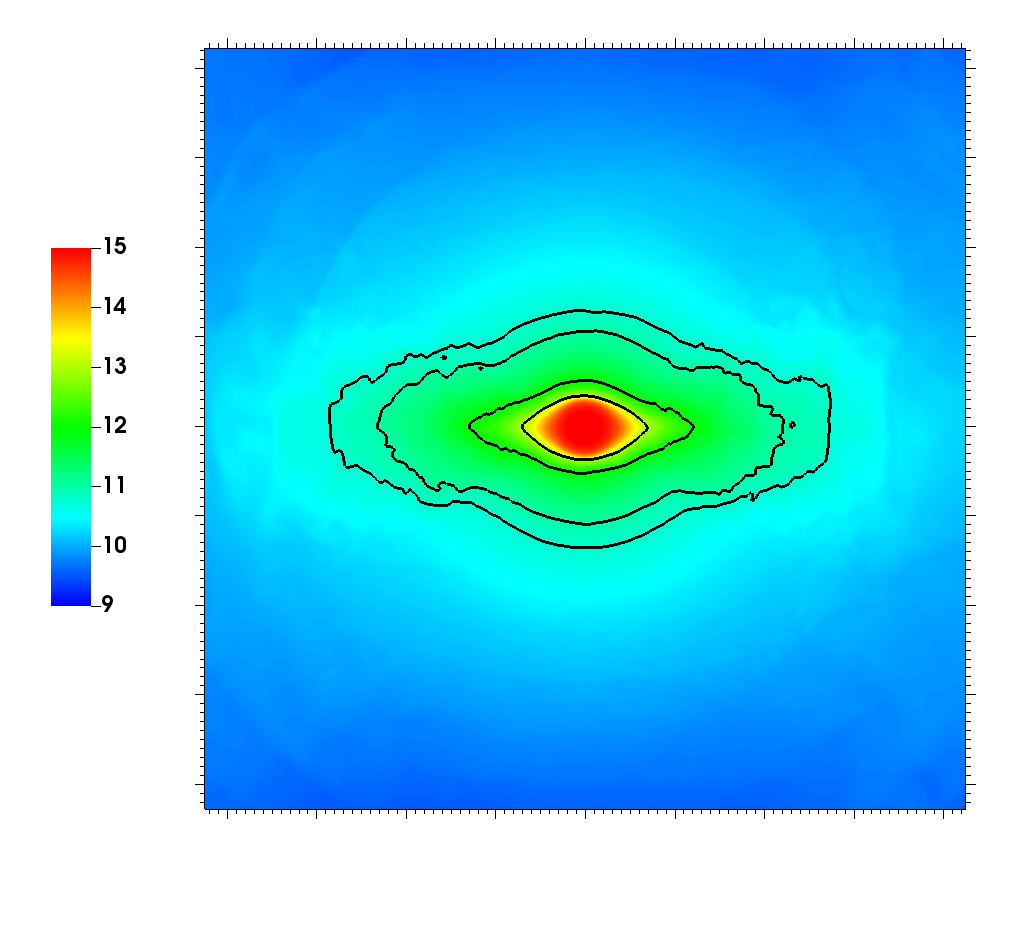}
	\includegraphics[width=0.3075\linewidth, trim={4.5cm 4.5cm 0 0}, clip]{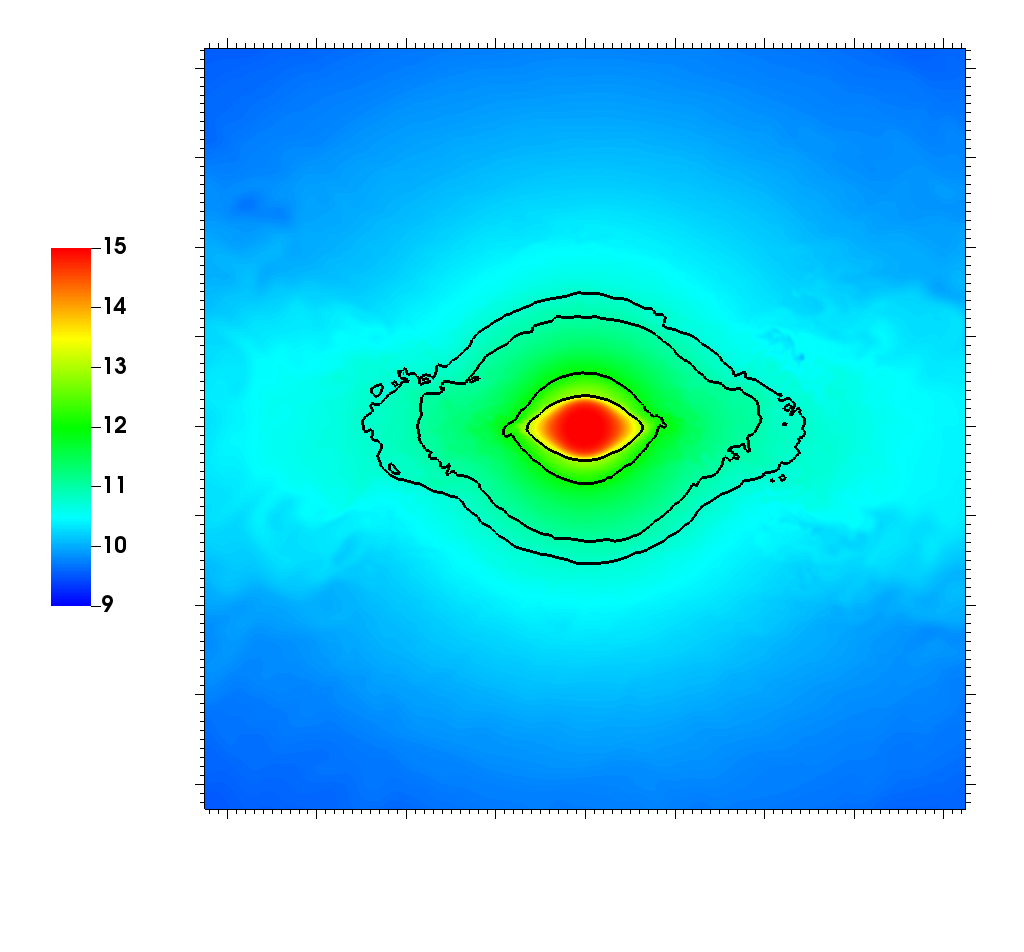}
	\includegraphics[width=0.35\linewidth, trim={0 2cm 0 0}, clip]{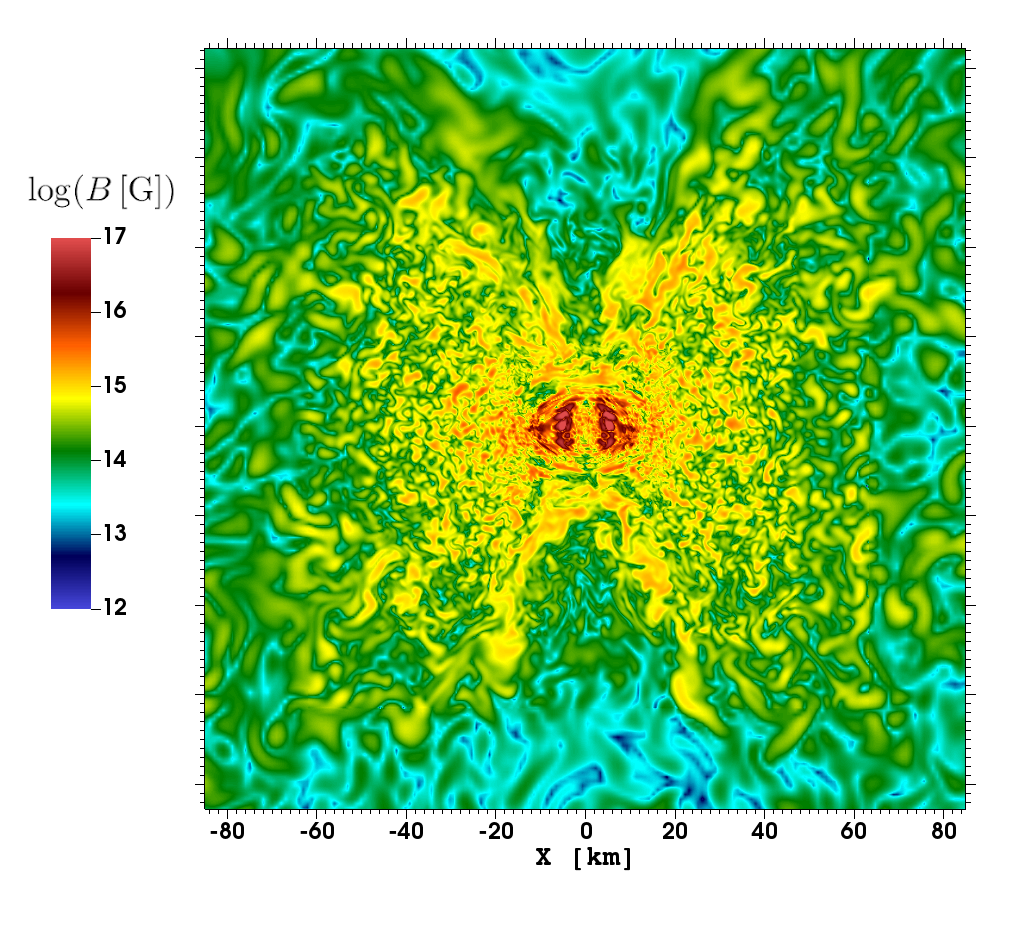}
	\includegraphics[width=0.3075\linewidth, trim={4.5cm 2cm 0 0}, clip]{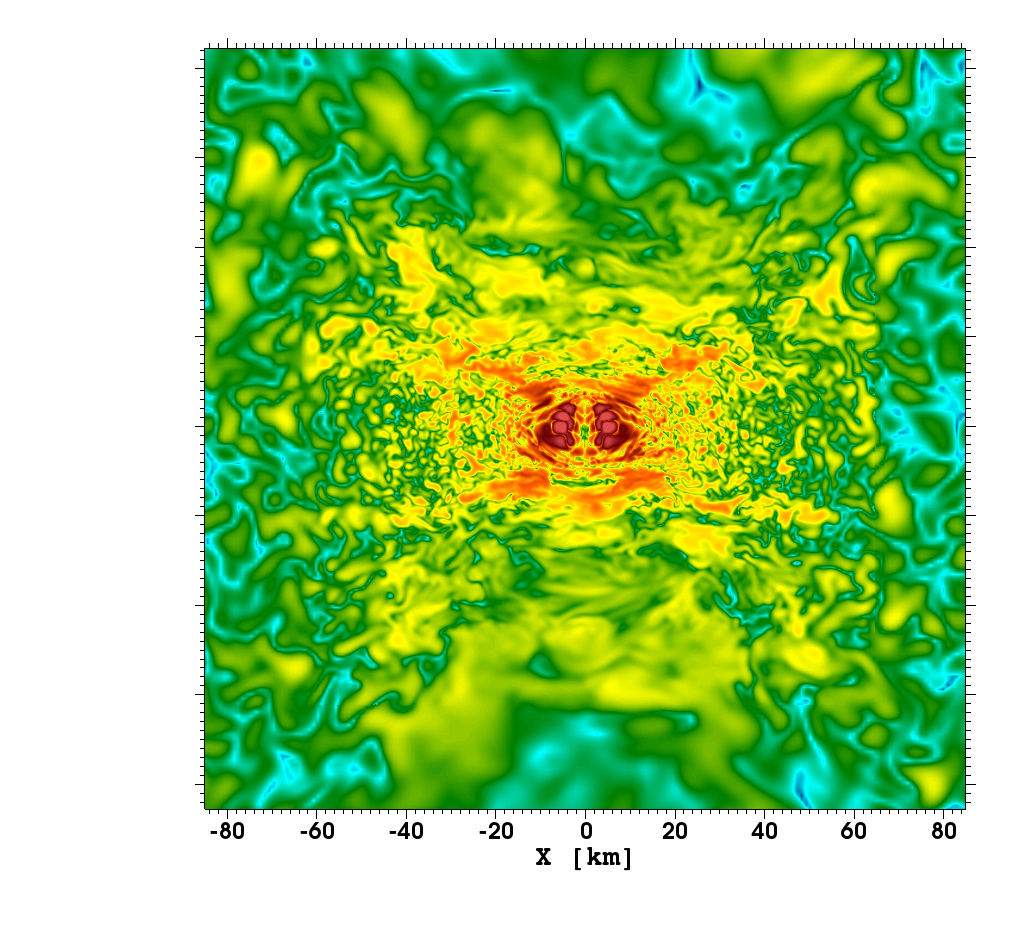}
	\includegraphics[width=0.3075\linewidth, trim={4.5cm 2cm 0 0}, clip]{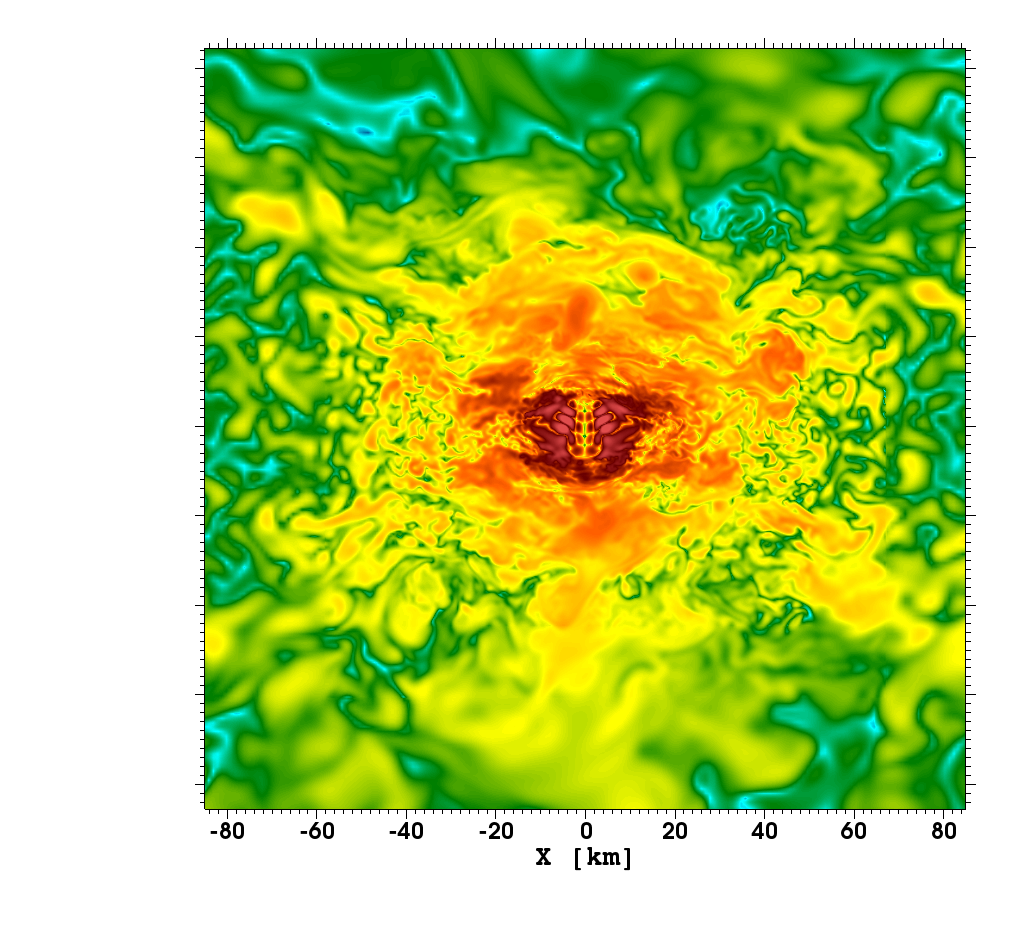}
	\caption{\textit{2D plots in the meridional plane}. (Top) Rest-mass density in [g~cm$^{-3}$]. The initial torus-like shape gets smoothed out as time progresses, becoming more spherical. In particular, the rotation axis is refilled with baryons, difficulting the formation of a magnetically dominated region. The solid black lines correspond to constant density surfaces, i.e. from the inner ones to outer, $\{10^{13}, 10^{12}, 10^{11}, 5 \times 10^{10}\}$~g cm$^{-3}$. Bottom: magnetic field intensity in [G]. The magnetic field strength grows continuously both in the bulk and the envelope, especially at late times. Columns represent different times after the merger, i.e. $t = \{50, 150, 250\}$ ms. The length is given in km.}
	\label{fig:2D_plots}
\end{figure*}

\begin{figure*}[ht!]
	\centering
	\includegraphics[scale=0.16]{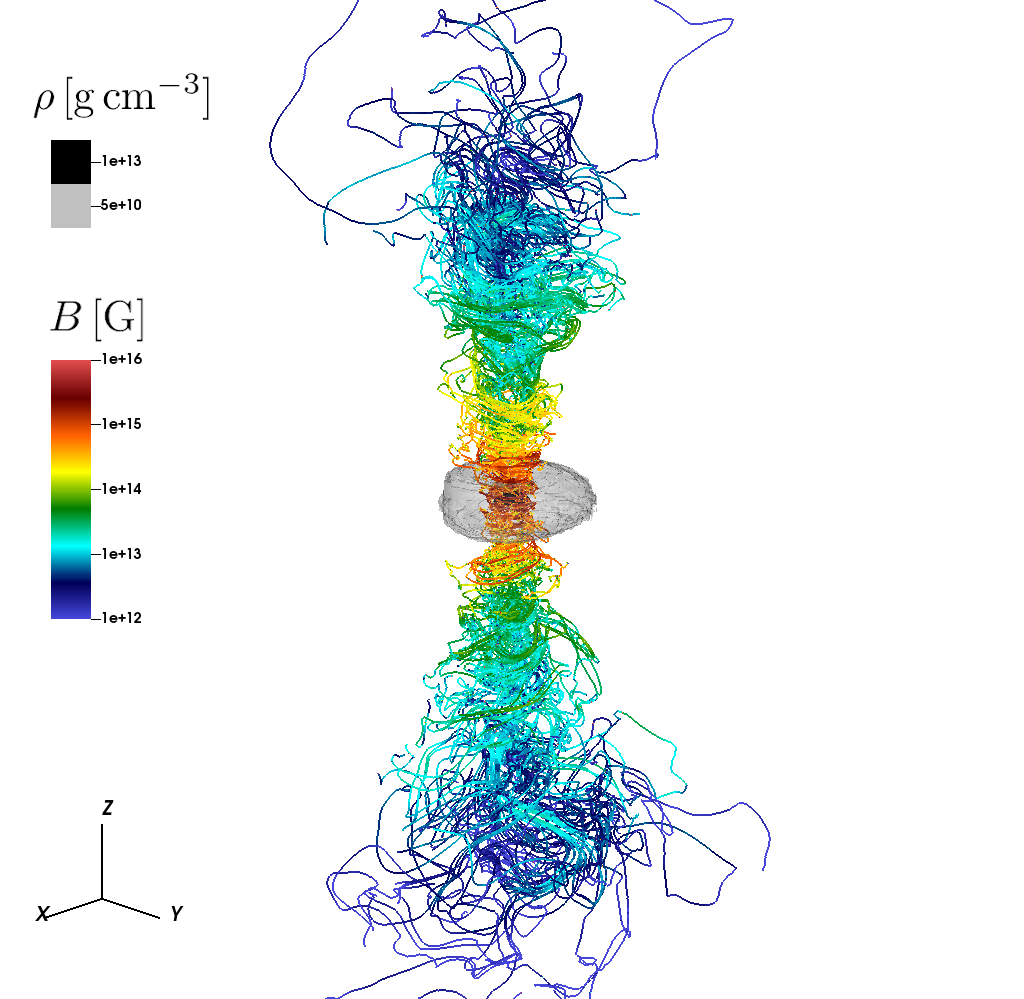}
	\includegraphics[scale=0.16]{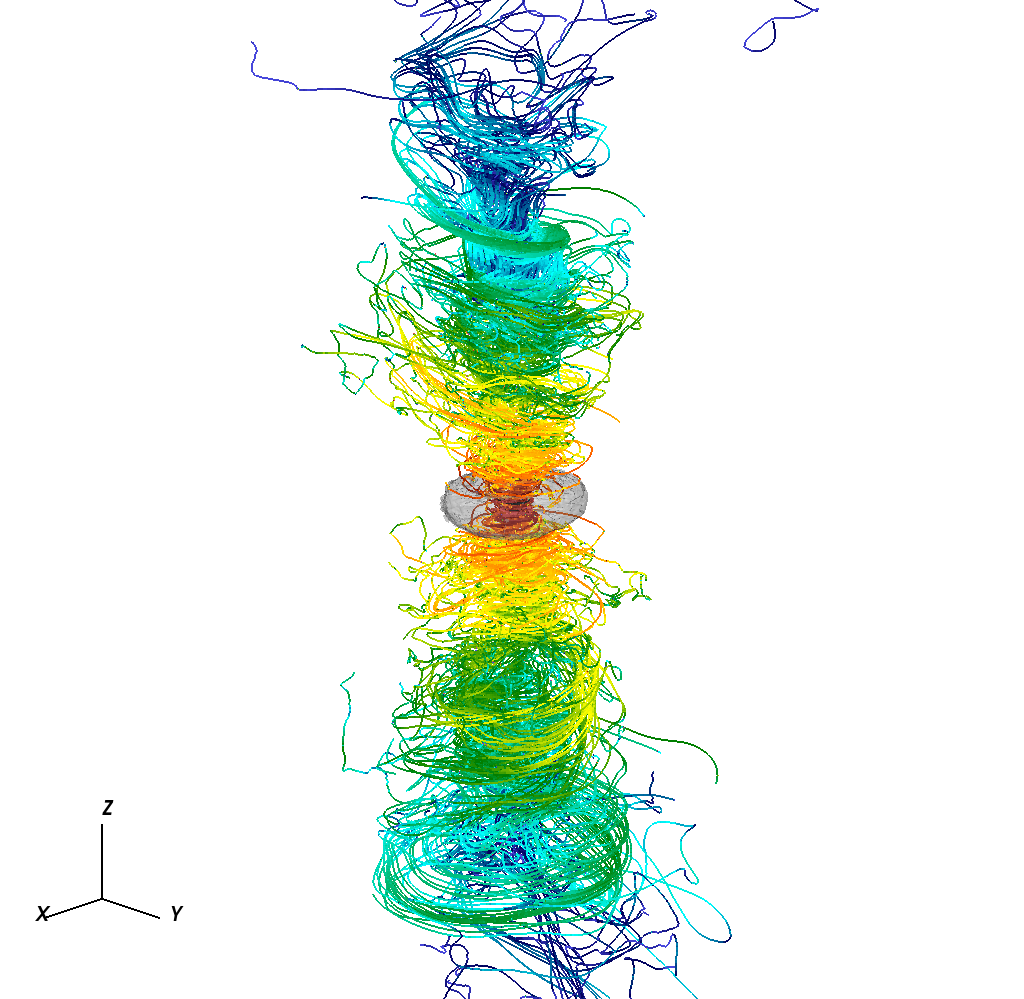}
	\includegraphics[scale=0.16]{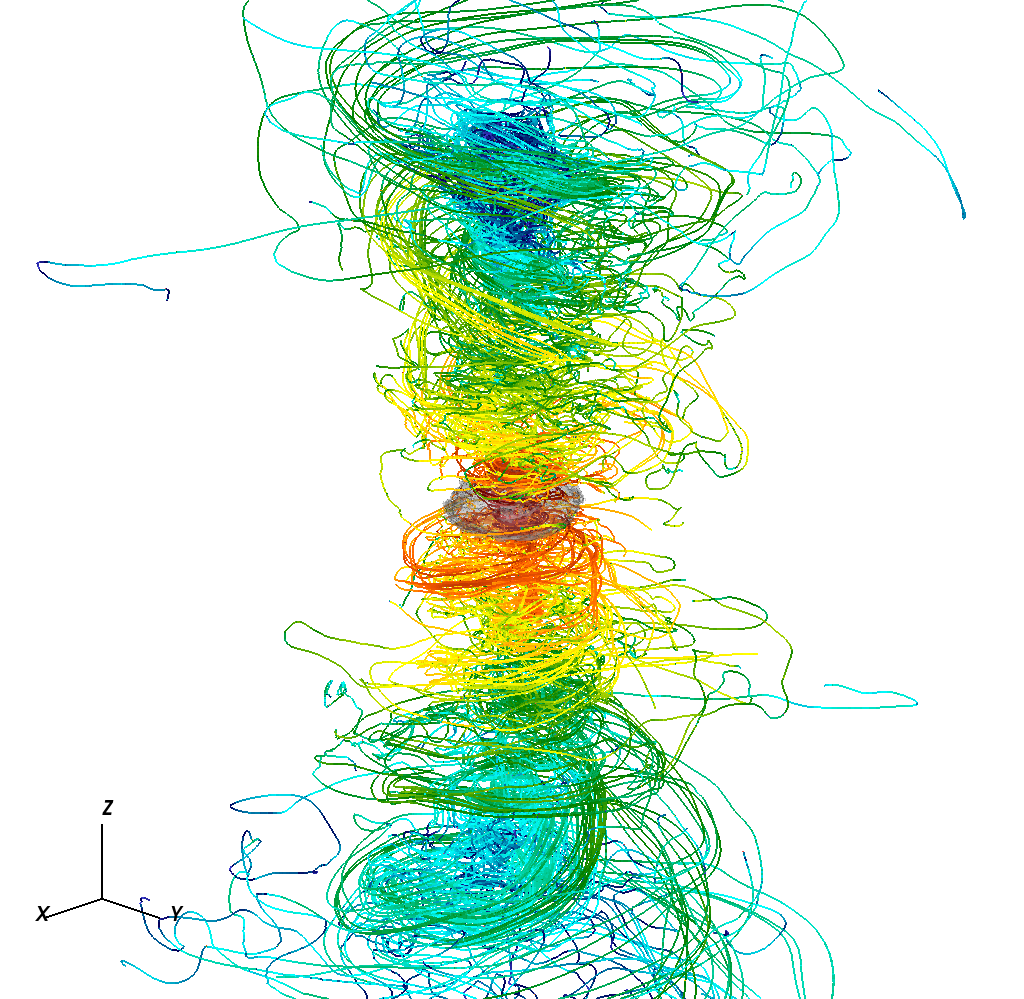}\\
	\caption{ \textit{Formation of large-scale magnetic field structures}. Streamlines of the magnetic field, at times $t = \{50, 150, 250\}$ ms after the merger.
		The density constant surfaces at $\rho = \{5\times10^{10}$, $10^{13}$\} g~cm$^{-3}$ are also shown. Notice the development of a helicoidal large-scale structure during the post-merger phase.}
	\label{fig:3D_streamlines}
\end{figure*}

\begin{figure}[ht!]
	\centering
	\includegraphics[width=0.9\linewidth]{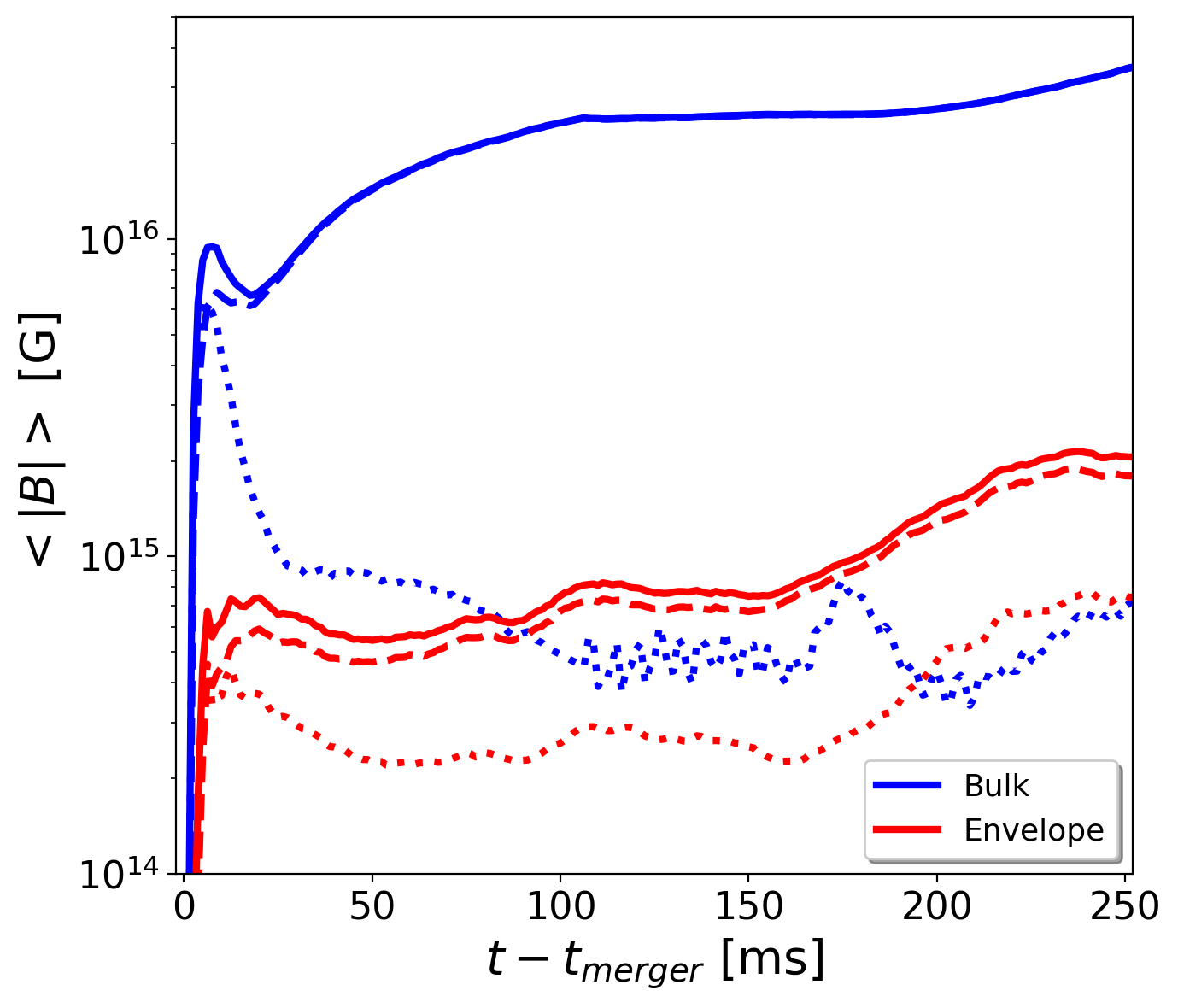}
	\caption{\textit{Average magnitude of magnetic field components}. Evolution of the averaged magnitude of the poloidal (dotted), the toroidal (dashed) and the total magnetic field (solid), both in the bulk (blue) and envelope (red) regions of the remnant. Notice both components of the magnetic fields in the envelope starts to grow significantly again from $t \sim 150$~ms on.}
	\label{fig:averagebbpolbtor_full}
\end{figure}

We use several integral quantities to monitor the dynamics in different regions: averages of the magnetic field poloidal $B_{pol}$ and toroidal $B_{tor}$ components, the fluid angular velocity $\Omega \equiv \frac{d\phi}{dt} = \frac{u^{\phi}}{u^t}$ (where $u_a \equiv W (-\alpha, v_i )$ denotes the fluid four-velocity), the plasma beta parameter, $\beta = \frac{2 P}{B^2}$ and the fastest growing unstable mode of the magneto-rotational instability (MRI)~\cite{balbus91}, $\lambda_{\textrm{MRI}} \approx \frac{B_{pol}}{\sqrt{4 \pi \rho}} \frac{2 \pi}{\Omega}$. The averages for a given quantity $q$ over a certain region ${\cal N}$ will be denoted generically by:
\begin{eqnarray}
{<}q{>}_a^b = \frac{\int_{{\cal N}} q \, d{\cal N}}{\int_{{\cal N}} d{\cal N}} ,
\end{eqnarray}
where ${\cal N}$ stands for a volume $V$ or a surface $S$, and the integration is restricted to regions where the mass density is within the range $(10^a, 10^b)\, {\rm g/cm^3}$. If $b$ is omitted, it means no upper density cut is applied. From here after we will denote the \textit{bulk} as the densest region of the remnant with densities $\rho > 10^{13}~g~cm^{-3}$, and the \textit{envelope}  as the region satisfying $5 \times 10^{10} ~g~cm^{-3} < \rho < 10^{13} ~g~cm^{-3}$. In particular, we define averages over the bulk of the remnant as ${<}q{>}_{13}$, and, abusing a bit of the notation for simplicity, averages over the envelope as ${<}q{>}^{13}_{10}$. Surface integrals are carried out over a cylinder $S$ with axis passing through the center of mass and orthogonal to the orbital plane. We also compute global quantities, integrated over the whole computational domain, such as the total magnetic energy $E_{mag}$, in order to monitor their evolution.

In addition, we compute the spectral distribution of the kinetic and magnetic energies over the spatial scales. For the magnetic spectra, we also calculate the poloidal and toroidal contributions separately. Further details of the numerical procedure to calculate the spectra can be found in~\cite{aguilera2020,vigano19,vigano20,palenzuela22}. With these spectral distributions we can define the spectrum-weighted average wave-number
\begin{equation}
\langle k \rangle \equiv \frac{\int_k k\,{\cal E}(k) \,dk} {\int_k {\cal E}(k)\, dk}~,
\end{equation}
with an associated length scale $\langle L \rangle = 2\pi/ \langle k \rangle $  which represents the typical \textit{coherent} scale of the structures present in the field.

Finally, we compute the non-axisymmetric intensity introduced in \cite{aguilera23}, a  topological indicator that quantifies the complexity of the field over the domain and can be a proxy to the amount of turbulence.
In particular, at each cylindrical radii $R$, it measures  the fraction of non-axisymmetric contributions to the total energy. Therefore, we can compute the non-axisymmetric intensity for both the magnetic and kinetic energy, namely \begin{equation}\label{eqn:Iturbulence}
	I_{kin}(R) = \frac{{\delta E}_{kin}}{{\bar E}_{kin} + {\delta E}_{kin}}
	~~,~~
	I_{mag}(R) = \frac{{\delta E}_{mag}}{{\bar E}_{mag} + {\delta E}_{mag}} ~.
\end{equation}
These axisymmetric indicators are zero for perfect axial symmetry and approach one if the non-axisymmetric modes are dominating. These modes can either have a turbulent origin (small scales), or consist of large/intermediate non-axisymmetric modes. Therefore, we infer how relevant the turbulence is by simultaneously looking at the axisymmetric indicators, the spectra and, visually, the orbital plane slices.

%%%%%%%%%%%%%%%%%%%%%%%%%%%%%%%%%%%%%
\section{Long evolution of the HMNS remnant} \label{sec:long}
%%%%%%%%%%%%%%%%%%%%%%%%%%%%%%%%%%%%%

In this section we summarize the analysis of the remnant produced by the BNS merger for our reference LES. The simulation presented here continues the one performed in~\cite{aguilera23}, evolving from $100$ to $250$~ms after the merger while decreasing the resolution from $60$ to $120$~m. We describe the qualitative dynamics, the energetics evolution and the spatial and spectral distribution of the HMNS remnant. In some of the plots we include the results from previous stages of the simulation for comparison purposes, as well as to find trends with respect to the new results.

%%%%%%%%%%%%%%%%%%%%%%%%%%%%%%%%%
\subsection{Qualitative dynamics and average fields}
%%%%%%%%%%%%%%%%%%%%%%%%%%%%%%%%%

In Fig.~\ref{fig:2D_plots} we display the density (top) and magnetic field strength (bottom) in a meridional plane at $t=\{50, 150, 250\}$ ms after the merger. 
%\RA{The solid black lines from the top panel represent constant density surfaces, i.e. from the inner lines to the outer ones, $10^{13}$, $10^{12}$, $10^{11}$, $5 \times 10^{10}$ g cm$^{-3}$}. 
Although initially the remnant has a torus-like shape, as time evolves the rotation decreases and the remnant becomes more spherical. The spatial distribution of the averaged magnetic field initially mimics the torus shape, but becomes more spherical as it builds up and gets amplified up to $10^{16}-10^{17}$~G in the bulk and in the inner part of the envelope.  We remind that the volume-averaged magnetic field in the bulk was amplified up to $10^{16}$~G during the first $5$~ms after the merger. After that, the magnetic field grows slowly, mainly in the toroidal component, suggesting that the amplification is sourced by winding during this stage.

\begin{figure*}[ht!]
	\centering
	\includegraphics[width=0.32\linewidth]{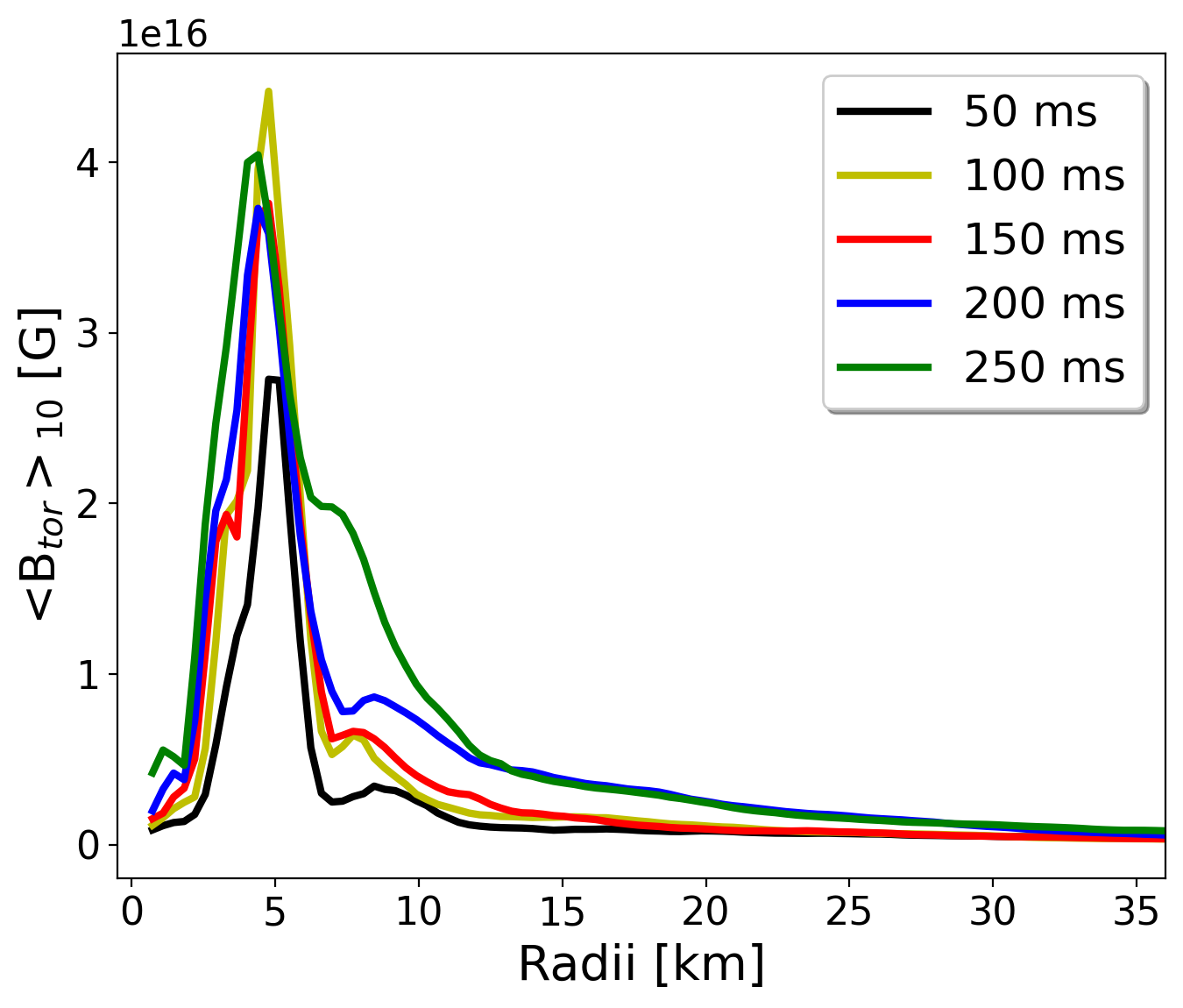}
	\includegraphics[width=0.32\linewidth]{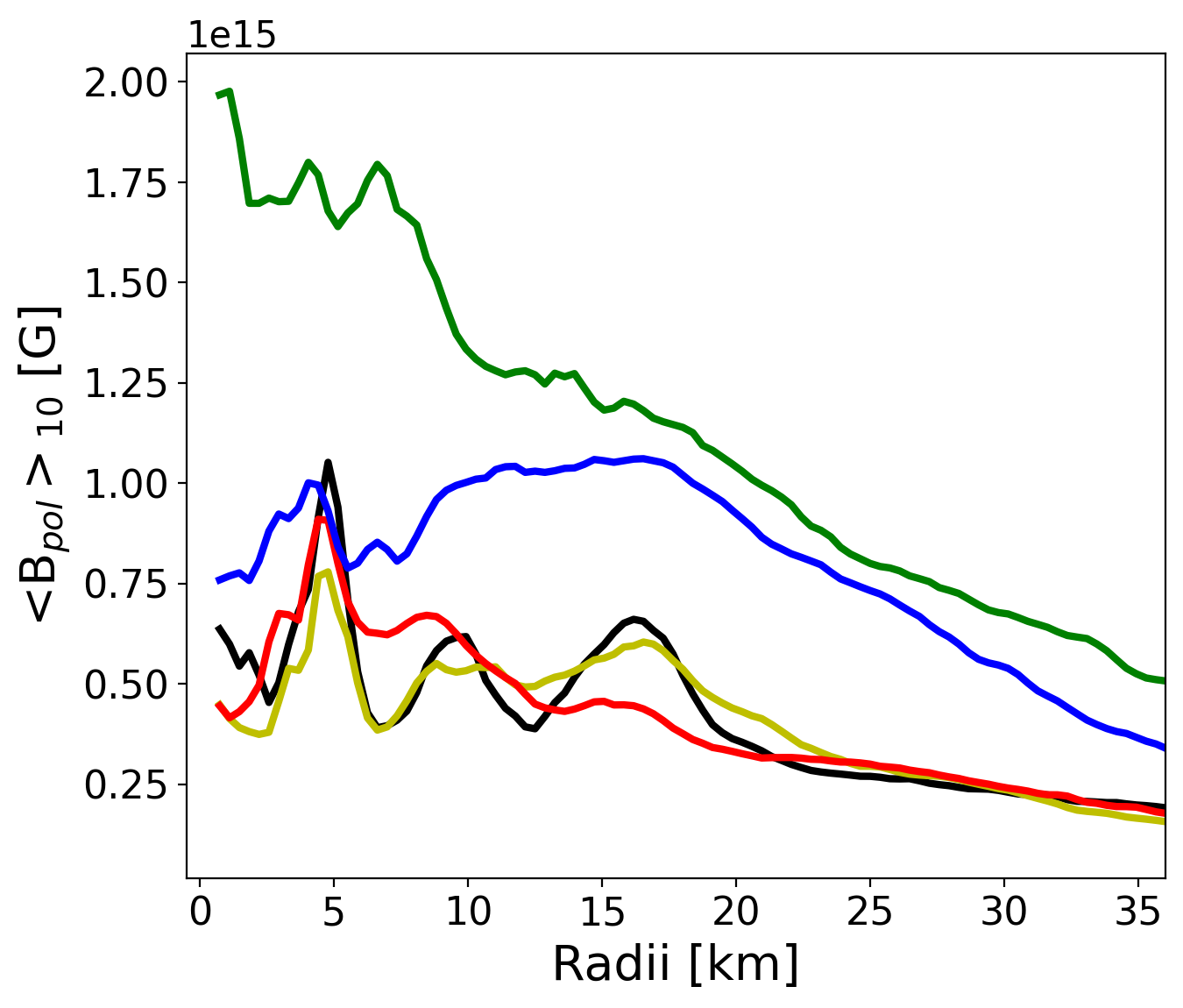}
	\includegraphics[width=0.32\linewidth]{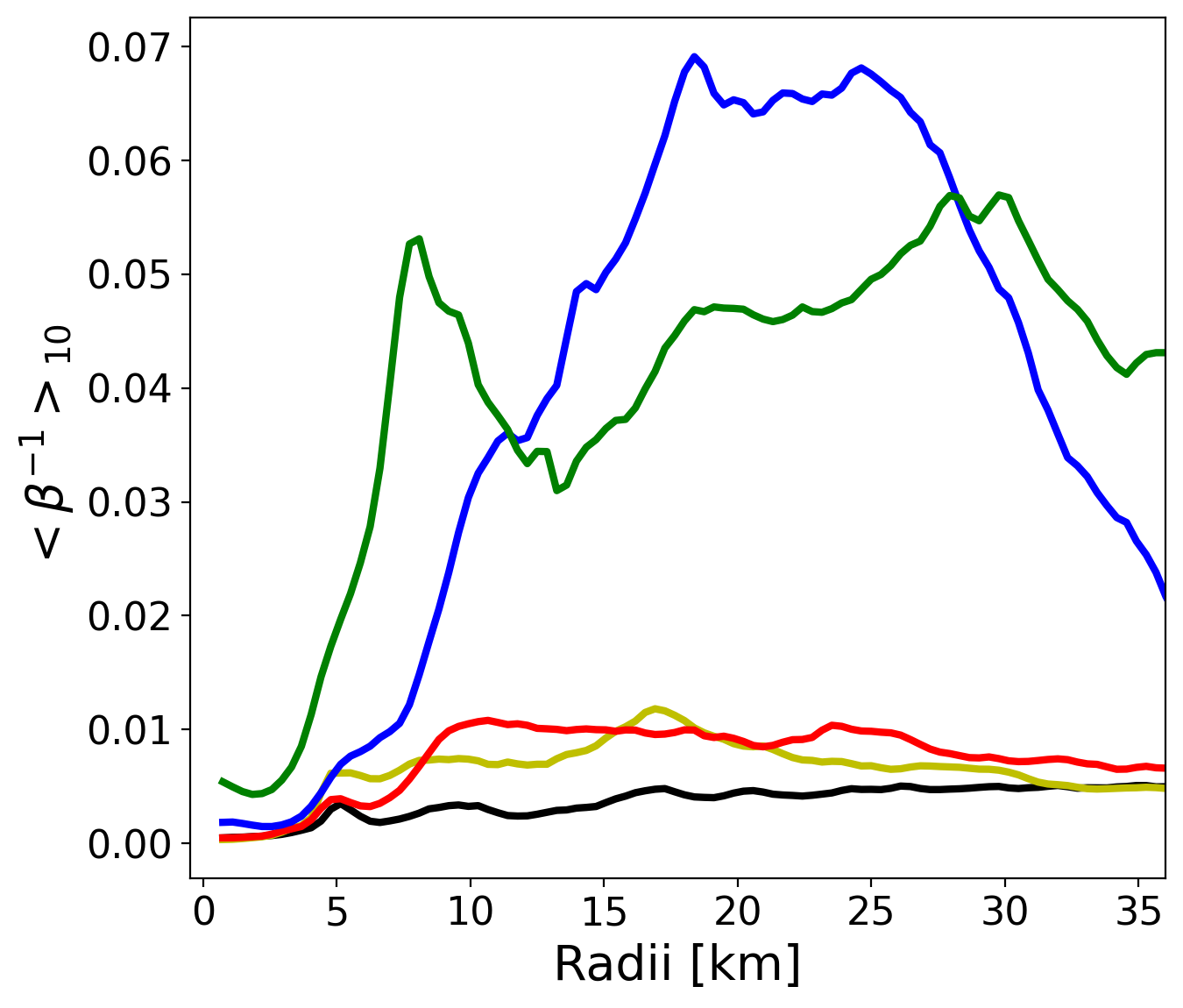}\\
	\includegraphics[width=0.32\linewidth]{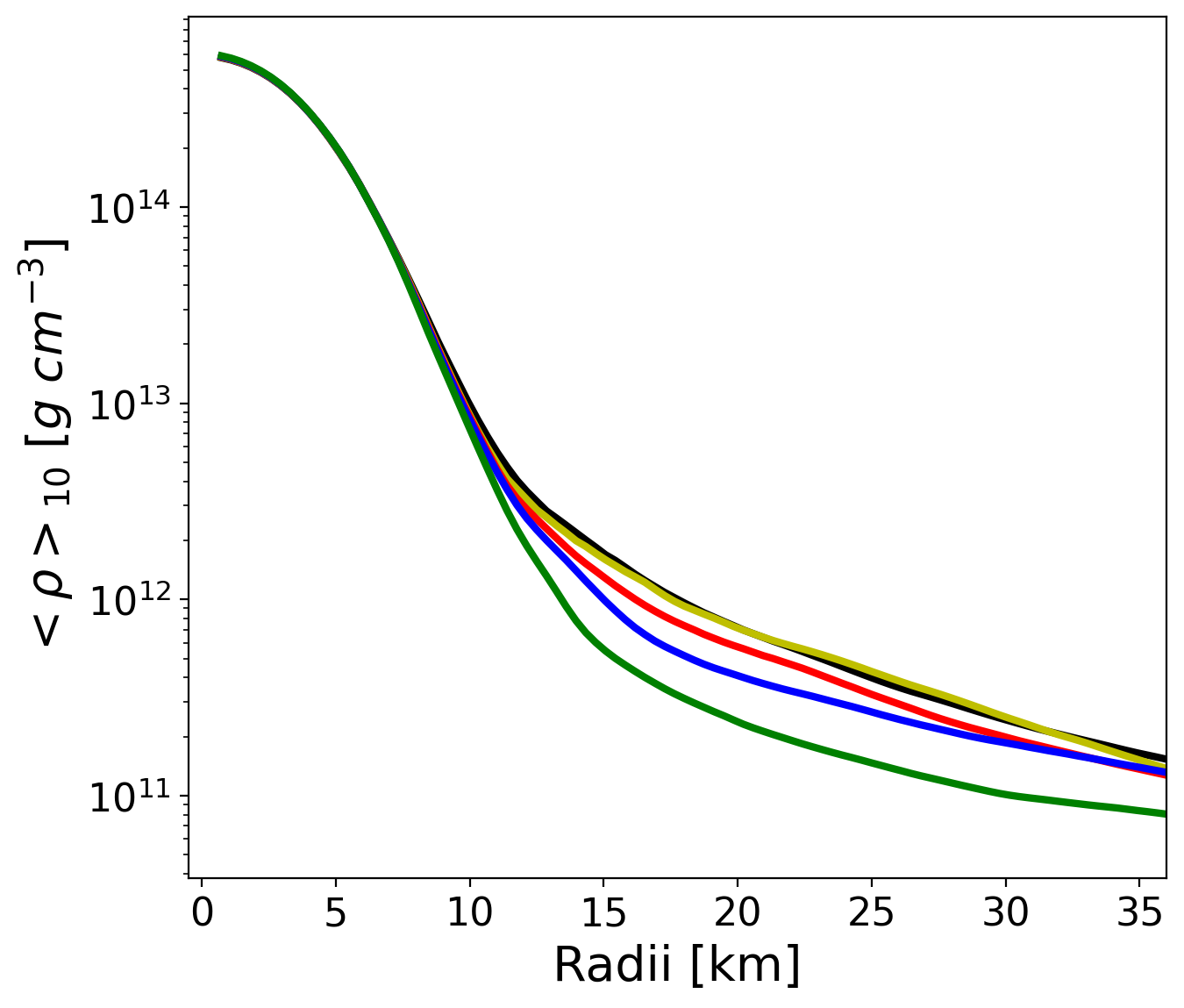}
	\includegraphics[width=0.32\linewidth]{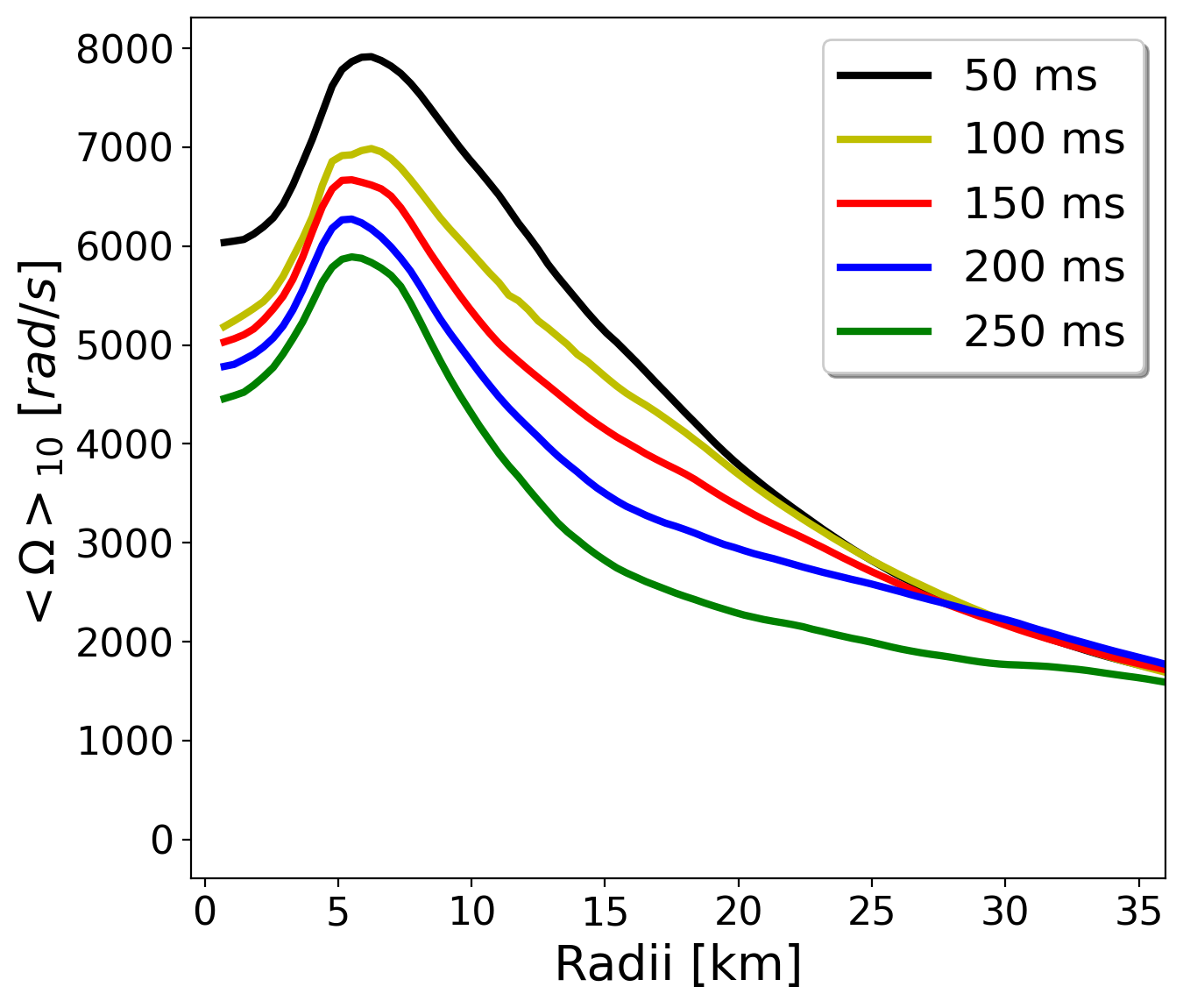}
	\includegraphics[width=0.32\linewidth]{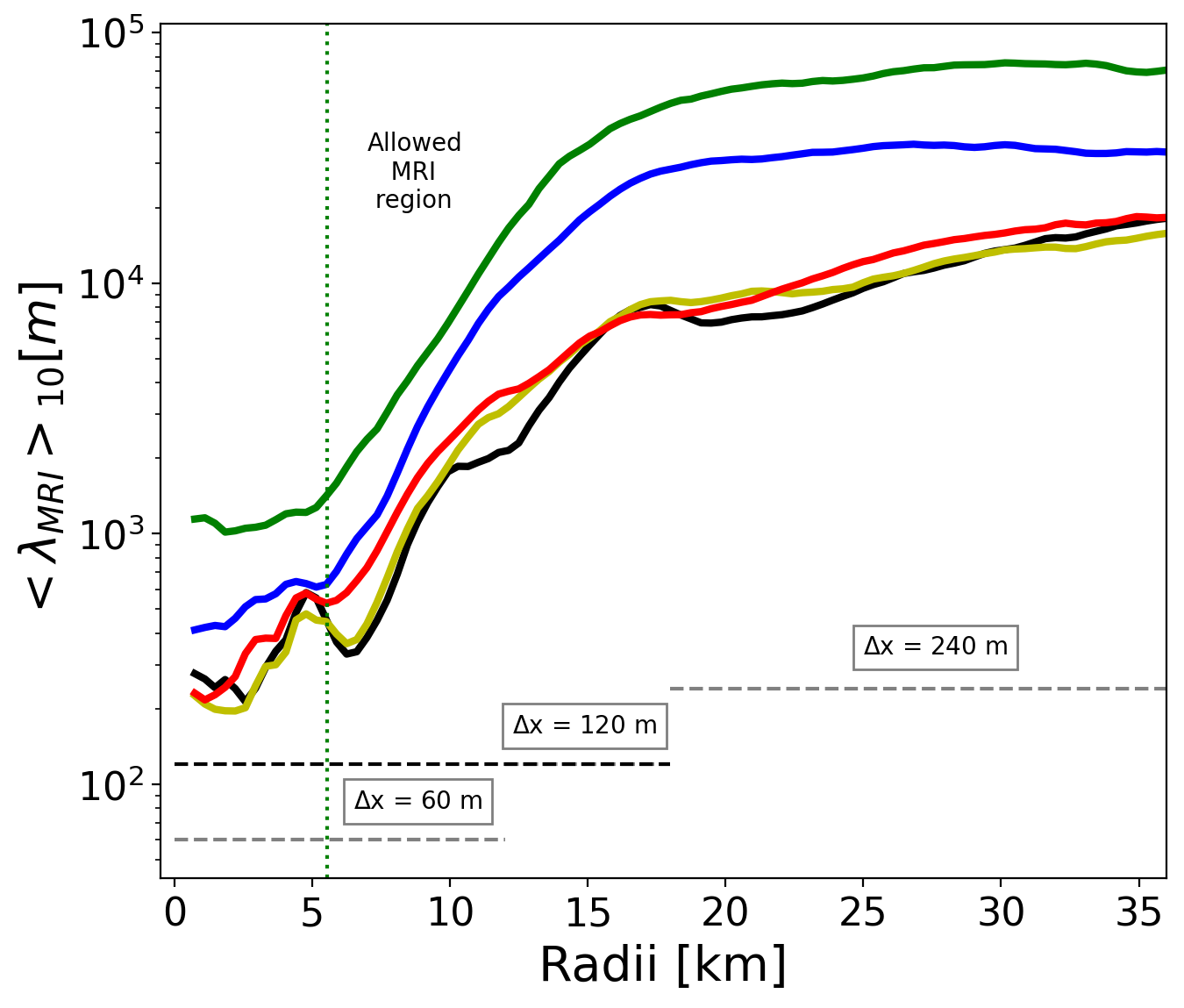}
	\caption{\textit{Spatial distribution evolution}. Average of different quantities as a function of the  cylindrical radius at different times. Notice that the poloidal component grows significantly at late times, especially at small radii. The (averaged) fastest unstable MRI mode, $\lambda_{MRI}$, is well captured with our grid resolution for $R \gtrsim 10$~km, and at all distances at late times $t\gtrsim 250$~ms.}
	\label{fig:cylinder_full}
\end{figure*}

\begin{figure}[ht!]
	\centering
	\includegraphics[width=\linewidth]{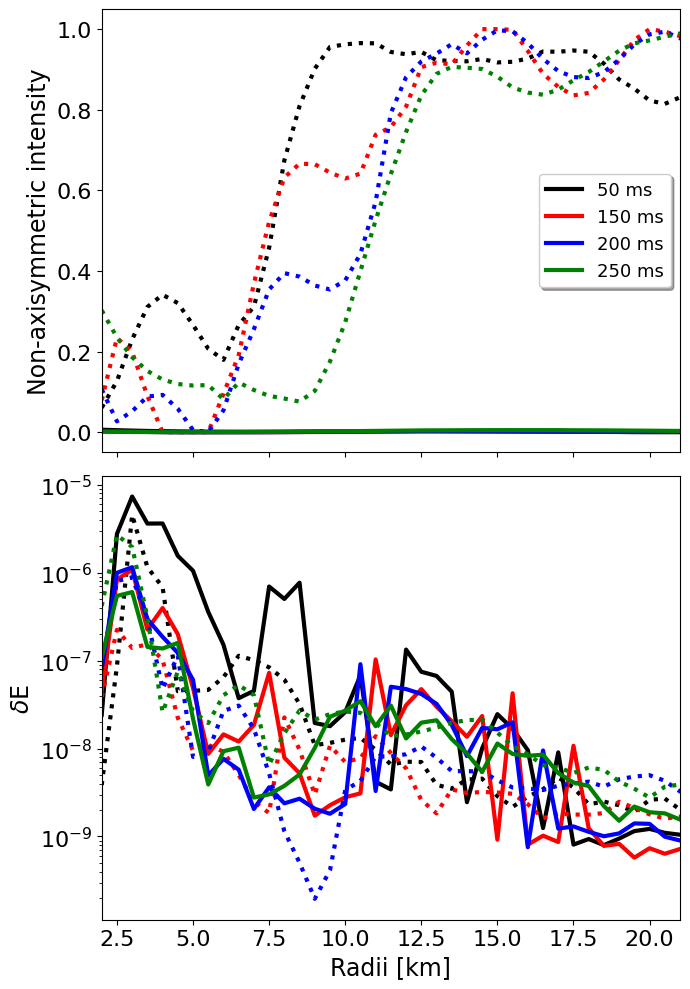}
	\caption{\textit{Estimates of the level of turbulence.} Kinetic (solid lines) and magnetic (dotted lines) non-axisymmetric intensities in the equatorial plane (top), together with its absolute values (bottom). The kinetic energy soon after the merger is always dominated by the azimuthal average of the velocity field. The magnetic field is slowly reaching a coherent axisymmetric structure in the bulk (i.e., up to $\sim 10$~km at $t = 250$~ms after the merger), but still presents high degree of turbulence in the envelope.  The quantities $\delta E_{kin}$ and $\delta E_{mag}$ are comparable both in the same time and radii, suggesting a strong connection between these non-axisymmetric contributions.}
	\label{fig:TI}
\end{figure}

The development of large-scale helicoidal structure can be observed more easily in the magnetic fieldlines, displayed in Fig.~\ref{fig:3D_streamlines}. Despite this clear structure of strong magnetic field strength (especially at late times), there is no magnetically dominated region with a stream of relativistic particles that could be associated to the launching of a jet.

%\RA{In Fig.~\ref{fig:Bynorm} it is shown the normalized poloidal magnetic field at different times after the merger, i.e. $t=\{200, 250, 300\}$ ms. There we can see that there is no clear structure of a jet to be launch at these times.}

The volume-averaged magnetic field strength is represented in Fig.~\ref{fig:averagebbpolbtor_full}. The blue colour represents the magnetic field in the bulk region of the remnant (from $10^{13}$ g cm$^{-3}$ and above) and the red one refers to the envelope (from $5\times10^{10}$ to $10^{13}$ g cm $^{-3}$). %The solid line correspond to the total magnetic field, and the dashed and dotted lines correspond to the toroidal and poloidal components, respectively. 
Both the bulk and the envelope are dominated by the toroidal components, at least up to $250$~ms after the merger. In the bulk, it continues monotonically growing, reaching  $B_{tor} \sim 4 \times 10^{16}$~G by the end of our simulation. Some oscillations are visible in the envelope, although there is an average growth, especially after $t \sim 150$~ms. The poloidal component in both regions have some oscillations from $100$~ms until the end of the simulation, but the one in the envelope shows a new significant monotonic growth starting approximately at $150$~ms. By the end of the run, the poloidal component in both regions have a comparable strength $B_{pol} \sim 10^{15}$~G.

%%%%%%%%%%%%%%%%%%%%%%
\subsection{Spatial distribution}
%%%%%%%%%%%%%%%%%%%%%%

We can now analyze the spatial distribution of our numerical solutions. Several fields, averaged on cylinders of height $30$~km, are displayed in Fig.~\ref{fig:cylinder_full}. The toroidal and poloidal components of the magnetic field, together with the inverse of the plasma beta parameter, $\beta^{-1}= P_{\rm mag}/P_{\rm gas}$, are displayed in the top row.  
The toroidal component reaches $4 \times 10^{16}$~G on a broad-spike %\DV{{\bf broad-spike? Maybe: the average toroidal component sharply peaks at 5 km with 4e16 G?}}
region approximately at $5$~km, with only minor, slow shifts during the evolution. The poloidal component shows a flatter profile with $\sim 5 \times 10^{14}$~G since early times, that also evolves slowly until $t = 150$~ms. Then it starts to increase rapidly both in the bulk and the envelope by factors of a few. This shift in tendency, indicating a change in the dynamics, is confirmed also by the ratio $\beta^{-1}$. In the bottom row, the density, the angular velocity and the MRI wavelength $\lambda_{\textrm{MRI}}$ are displayed. The density changes are noticeable only in the envelope, while that the angular velocity decreases slowly overall. The fastest MRI-unstable mode is very large in the envelope, but reduces to hundreds of meters in the bulk mainly due to the high densities. We also display the grid resolution used in each region as to estimate roughly where these modes could be captured in our simulations. Although the final answer will also depend on the numerical methods employed, it appears that, on average, there should be enough resolution  to resolve most of the allowed MRI region where $\partial_R \Omega < 0$.

In Fig.~\ref{fig:TI} we plot the non-axisymmetric intensity as a function of radius in the top panel, while $\delta E_{kin}$ and $\delta E_{mag}$ are displayed in the bottom one. These quantities convey information about the degree of turbulence in the kinetic and magnetic fields (solid and dotted lines, respectively). In~\cite{aguilera23} it was shown that from the very beginning, the kinetic intensity was highly axisymmetric ($I_{kin} \sim 10\%$), with values close to $0$ for all radii. This means that there is only a small degree of (relative) turbulence on the velocity fields. As expected, at $250$ ms after the merger, it is even less turbulent and more axisymmetric.
%{\bf [Is it possible to say how much this is purely numerical in origin? ]}. 
On the other hand, the magnetic intensity showed a clear non-axisymmetric structure at the very beginning that was slowly shifted to an axisymmetric shape in the bulk (i.e., from $I_{mag} \sim 100\%$ to $I_{mag} \sim 10\%$). At $t = 250$~ms after the merger, we can observe in the non-axisymmetric indicator that the magnetic field is almost axially symmetric up to $10$~km, but it presents still high non-axisymmetric contributions at larger radii. Interestingly, the quantities $\delta E_{kin}$ and $\delta E_{mag}$ have comparable values (i.e., same order of magnitude) both in time and radii, suggesting that the kinetic and magnetic turbulent states are strongly connected. 
%Wether this correlation is a consequence of the MRI being active in the envelope of just a consequence of a decaying %turbulence scenario is not clear and require a much more refined analysis and resolution studies which will be addressed in the future.
Whether this correlation is due to the MRI being active within the envelope or a result of decaying turbulence remains unclear. Resolving this ambiguity likely requires further resolution studies and more refined analyses, which will be addressed in future research.
%\DV{{\bf This part needs a furtehr explanation, I don't touch it, beacuse I am not sure we all agree. My opinion: from the absolute values one simply infers that the turbulence is slowly decaying, no extra energy seems to be injected or excited, so no indication again for MRI, simply decaying turbulence. Basically, the axisymmetric kinetic energy is kept much longer because of the rotation (while the axisymmetric magnetic energy has built up from small scales), so in the top panel the relative intensity for kinetic (almost 0) and magnetic (approaching 1) are very different; in the bottom panel, the results are not biased by the inclusion of the big rotational budget, and curves show a normal, decaying turbulence. The question to be answered would be: why the axisymmetric magnetic field decays faster than the non-axisymmetric one (while for the kinetic energy it's possibly not true). In any case, in the top panel the kinetic energy lines have no information, they are too close to zero... Maybe a log scale?}}.

%%%%%%%%%%%%%%%%%%%%%%%%%%%%%%%%%%
\subsection{Spectral distribution}
%%%%%%%%%%%%%%%%%%%%%%%%%%%%%%%%%%

\begin{figure*}
	\centering
	\includegraphics[width=\linewidth]{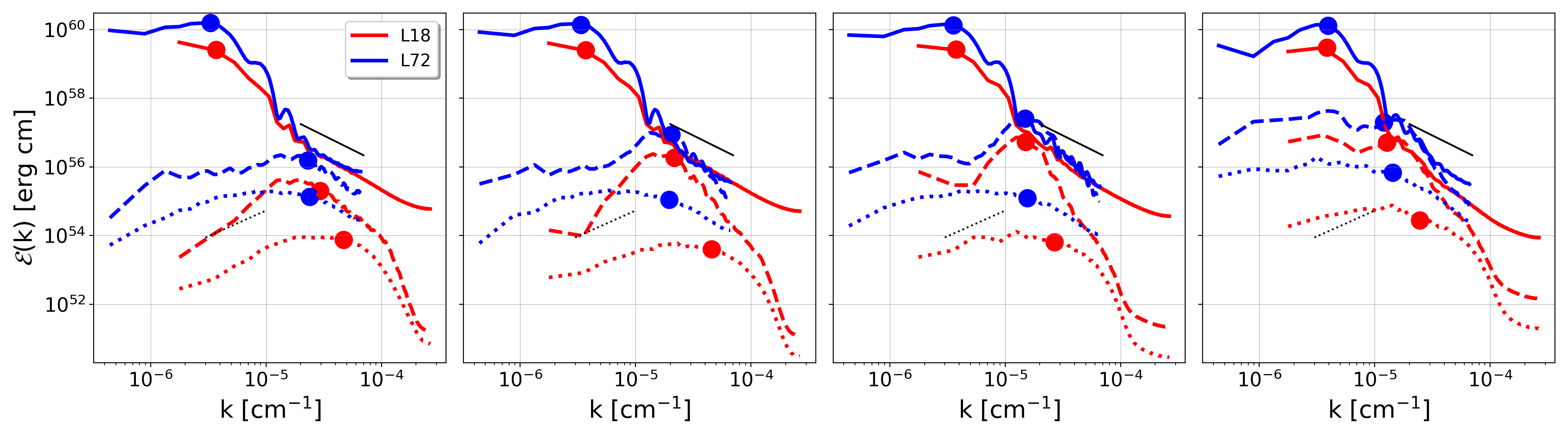}
	\caption{\textit{Spectral energy evolution}. Kinetic (solid), toroidal (dashed) and poloidal (dotted) magnetic energy spectra as a function of the wavenumber at $t = \{25,50,150,250\}$ ms after the merger. Different colors represent different box domains employed in the calculation, keeping the same number of points (i.e., decreasing the spatial resolution), such that the small domain (red) represents roughly the bulk and the large domain (blue) the bulk and the envelope.}
	\label{fig:spectra_full}
\end{figure*}

\begin{figure}
	\centering
	\includegraphics[width=\linewidth]{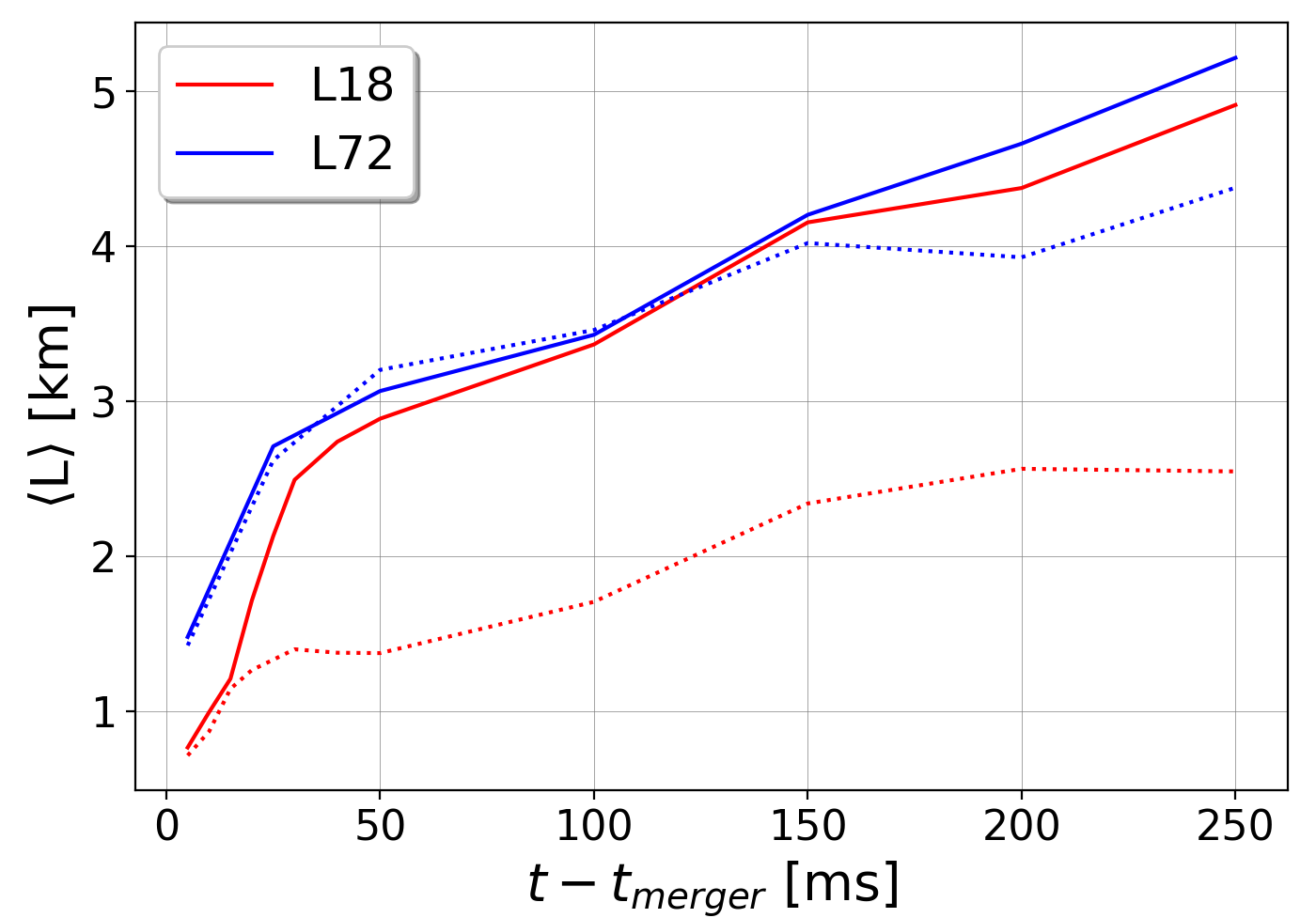}
	\caption{\textit{Evolution of the average characteristic lengthscale}. Average characteristic length of toroidal (solid) and poloidal (dotted) components of the magnetic field as a function of time for the different box domains (i.e., in the bulk in red and in the bulk plus envelope in blue), computed using $\langle L \rangle = 2\pi/ \langle k \rangle $. Notice that the characteristic length associated to the toroidal components is comparable in both regions. However, the poloidal one in the bulk is much smaller.}
	\label{fig:Average_characteristic_wavelength_full}
\end{figure}

The kinetic and magnetic energy spectra 
%(kinetic energy in solid lines, dashed line correspond to the toroidal component of the magnetic energy and the dotted ones represent the poloidal component)
displayed in Fig.~\ref{fig:spectra_full} are computed at $t=\{25, 50, 150, 250\}$ ms after the merger by considering two different regions: (i) a domain of $18$~km (L18, red colour) covering the bulk and the inner part of the envelope, and (ii) a domain of $72$~km (L72, blue colour) which encompasses the bulk and part of the envelope. The latter, due to its extended domain, could only be computed with a relatively low resolution $\Delta x = 480$~m, such that only the large-scale features are captured in the analysis. %The dotted and solid short straight lines correspond to Kazantsev and Kolmogorov slopes, $k^{3/2}$ and $k^{-5/3}$ respectively. 
As shown in \cite{aguilera23}, the toroidal magnetic energy reaches equipartition with the kinetic one within the bulk at length scales $\sim 3$~km in less than $50$~ms after the merger. The poloidal component remains smaller for all the timespan of the simulation, although it grows significantly after $150$~ms, especially at low wavenumbers (i.e., large spatial scales). The fact that both components grow at large scales starting from that time, confirms the trend shift mentioned before.

The evolution of the characteristic scales (i.e., the thick points that appear in the toroidal and the poloidal components at each time) is shown in Fig.~\ref{fig:Average_characteristic_wavelength_full}. The length-scale of the toroidal component in both regions  grows rapidly during the first $30$~ms after the merger, and then is approximately linear up to $250$~ms. The poloidal component also shows a similar behaviour, but while the one computed in the large domain is comparable to the toroidal component, the one in the small domain is approximately 3 times smaller. This confirms again that most of this energy has a turbulent origin and is confined in very small scales. After $150$~ms, the poloidal component in the small domain saturates and does not change significantly up to the end of the simulation.

%%%%%%%%%%%%%%%%%%%%%%%%%%%%%%%%%%%%%%%%%%%%%
\section{Impact of the initial magnetic field in under-resolved simulations} \label{sec:impact}
%%%%%%%%%%%%%%%%%%%%%%%%%%%%%%%%%%%%%%%%%%%%%

\begin{figure}[h!]
	\centering
	\includegraphics[width=0.9\linewidth]{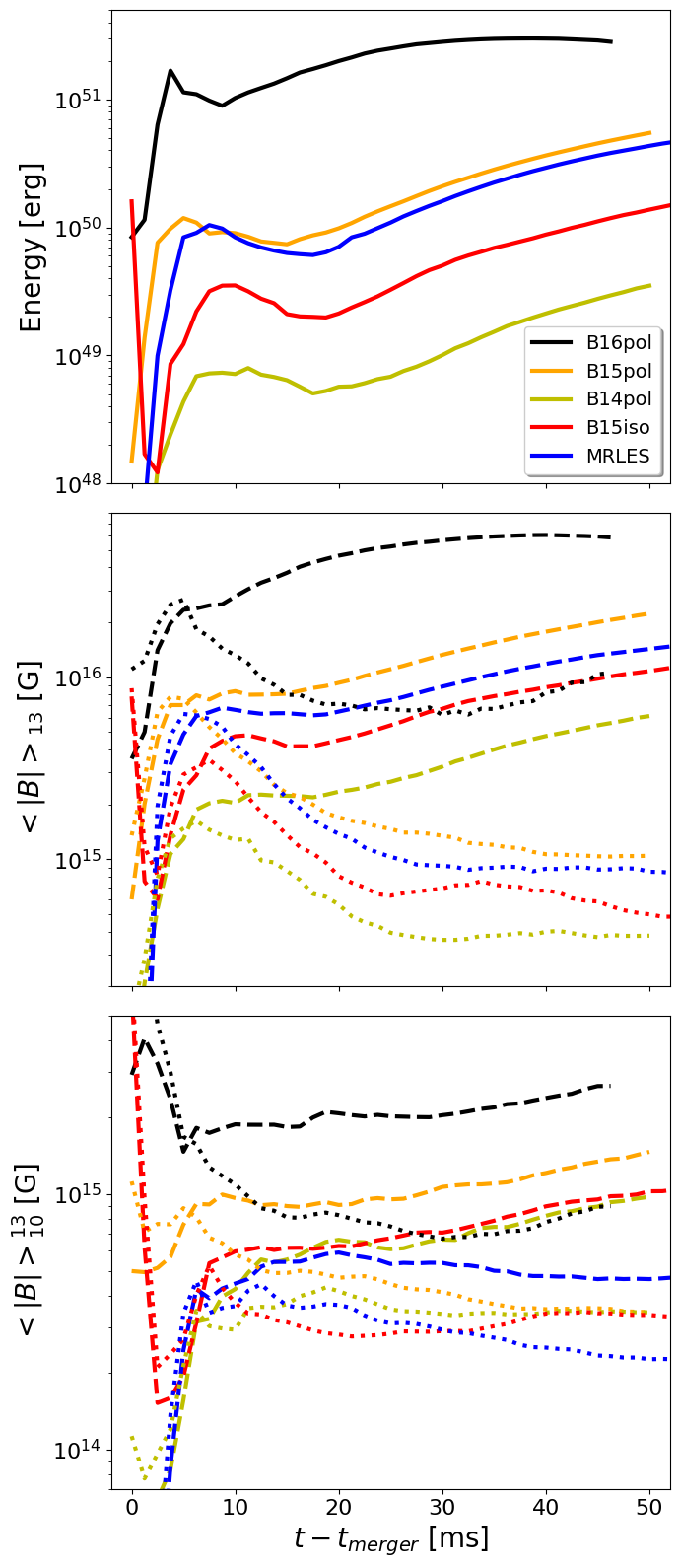}
	\caption{\textit{Global quantities for different simulations}. Evolution of the magnetic energy (top), and the averaged magnitude of the poloidal (dotted) and the toroidal (dashed) magnetic field components, both in the bulk (middle) and in the envelope (bottom) regions of the remnant, for the different simulations presented in Table \ref{tab:models}. }		
	\label{fig:integrals}
\end{figure}

\begin{figure*}[ht!]
	\centering
	\includegraphics[width=0.35\linewidth, trim={0 4.5cm 0 0}, clip]{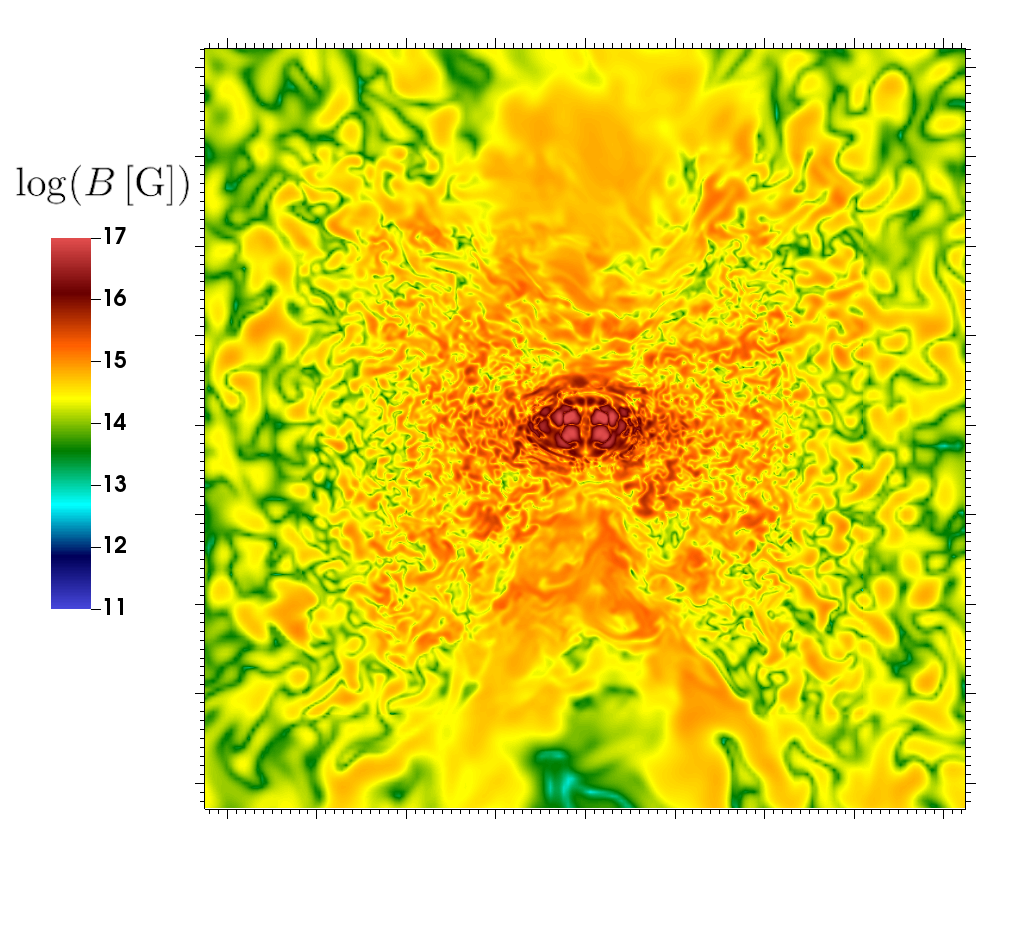}
	\includegraphics[width=0.3075\linewidth, trim={4.5cm 4.5cm 0 0}, clip]{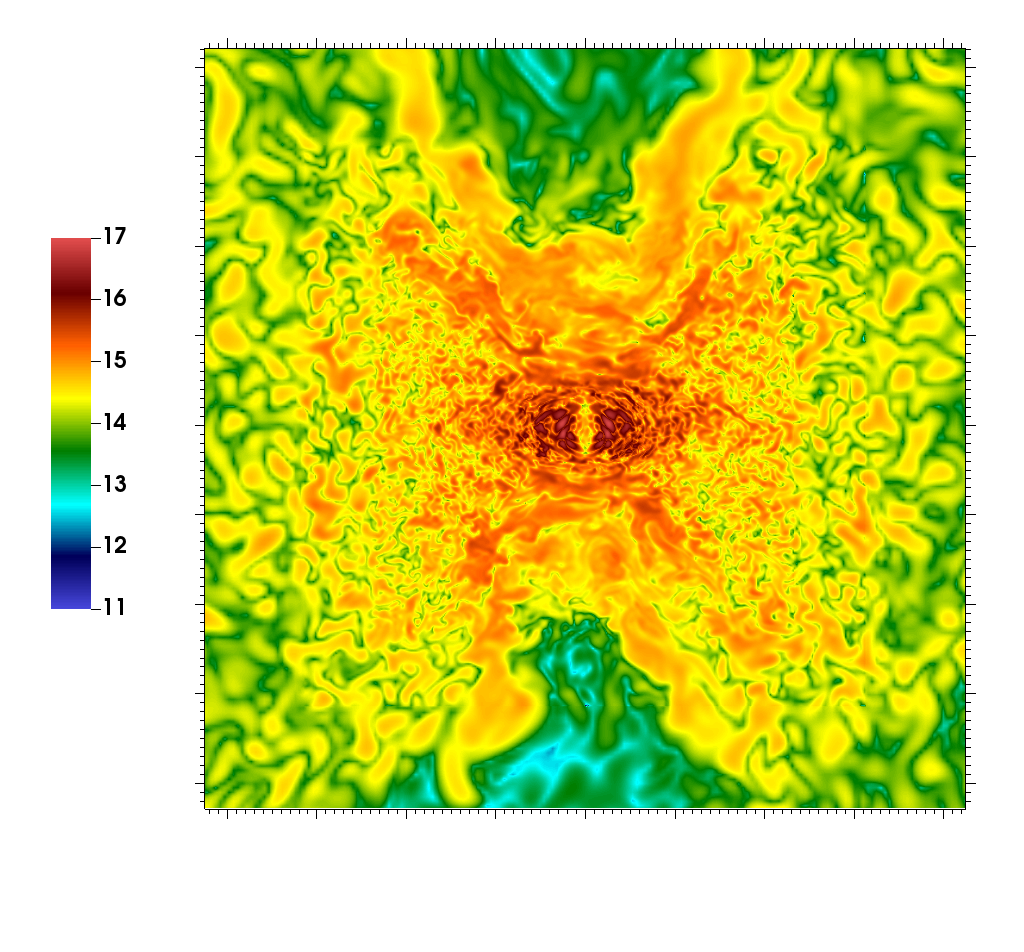}
	\includegraphics[width=0.3075\linewidth, trim={4.5cm 4.5cm 0 0}, clip]{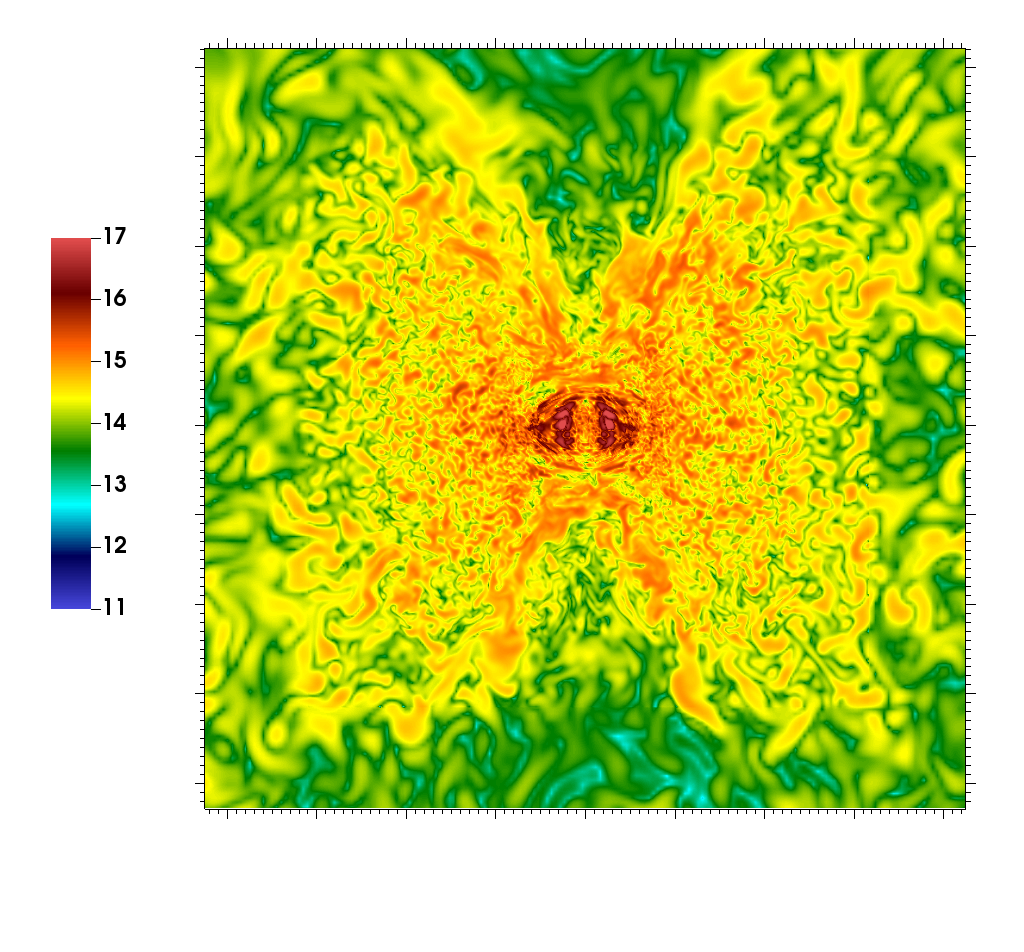}
	\includegraphics[width=0.35\linewidth, trim={0 2cm 0 0}, clip]{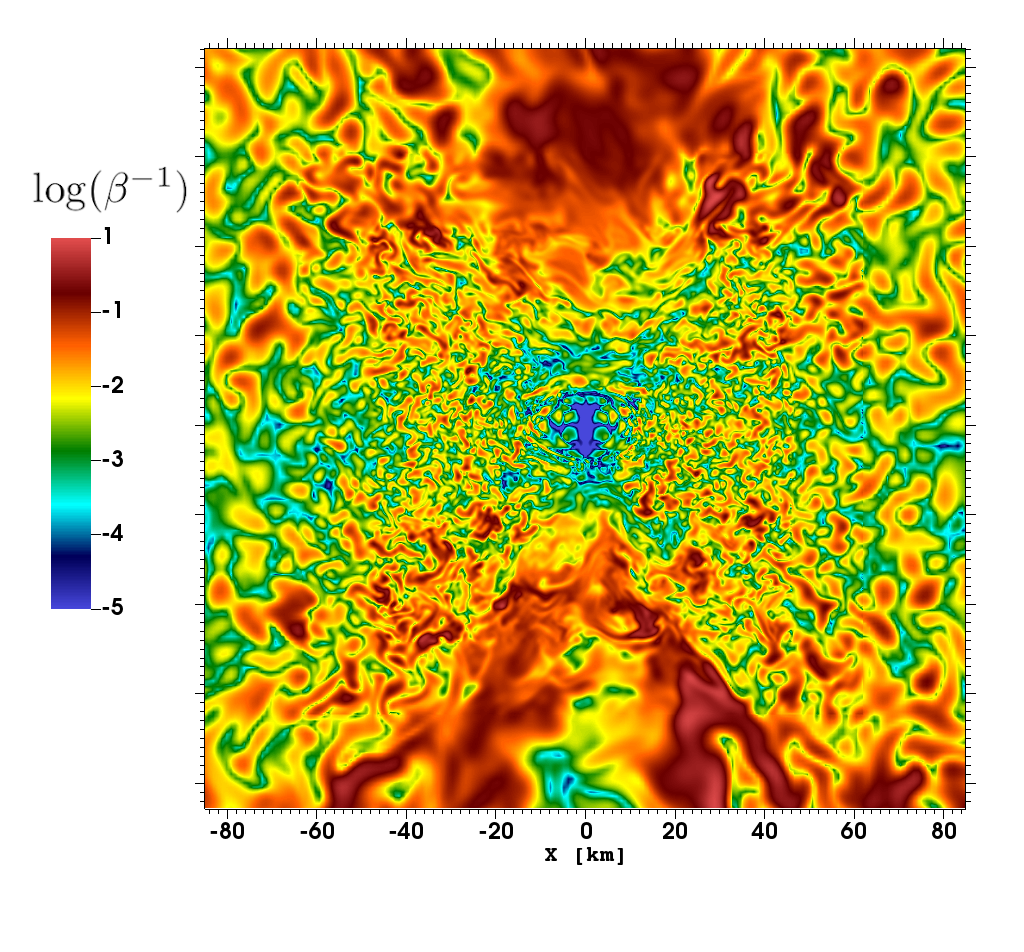}	
	\includegraphics[width=0.3075\linewidth, trim={4.5cm 2cm 0 0}, clip]{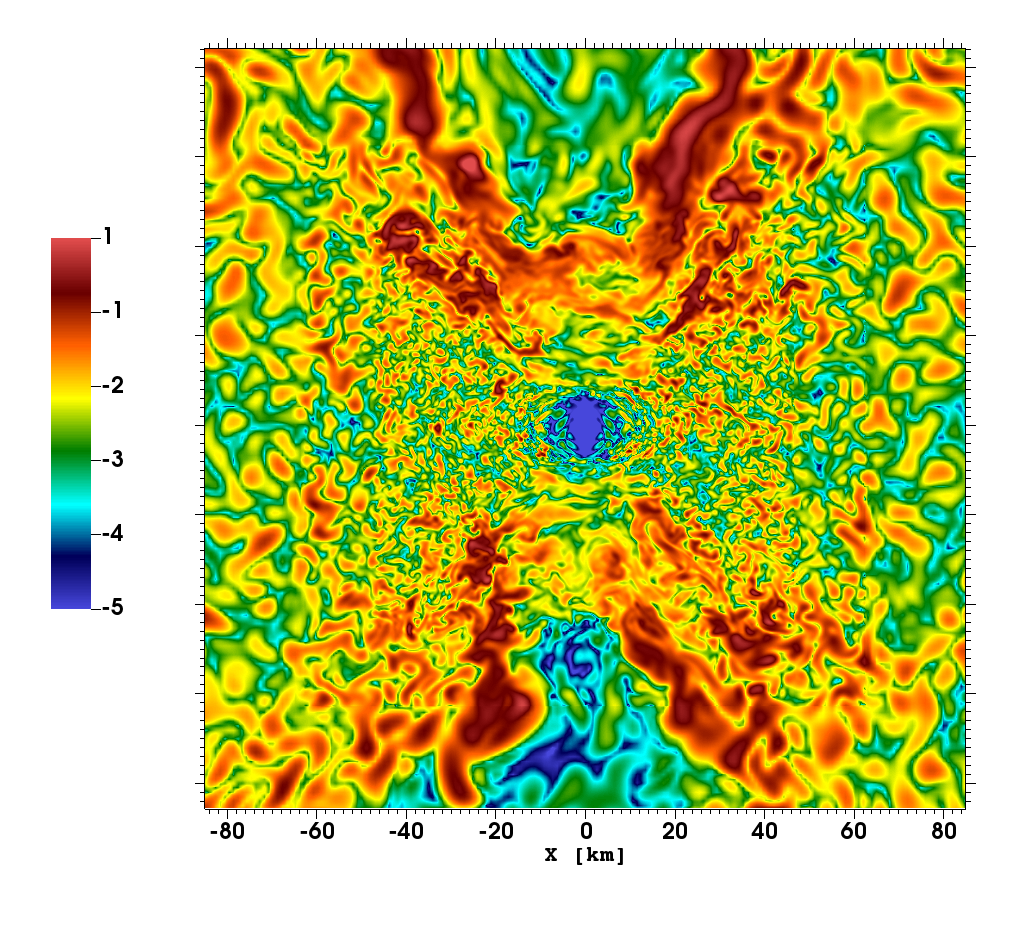}
	\includegraphics[width=0.3075\linewidth, trim={4.5cm 2cm 0 0}, clip]{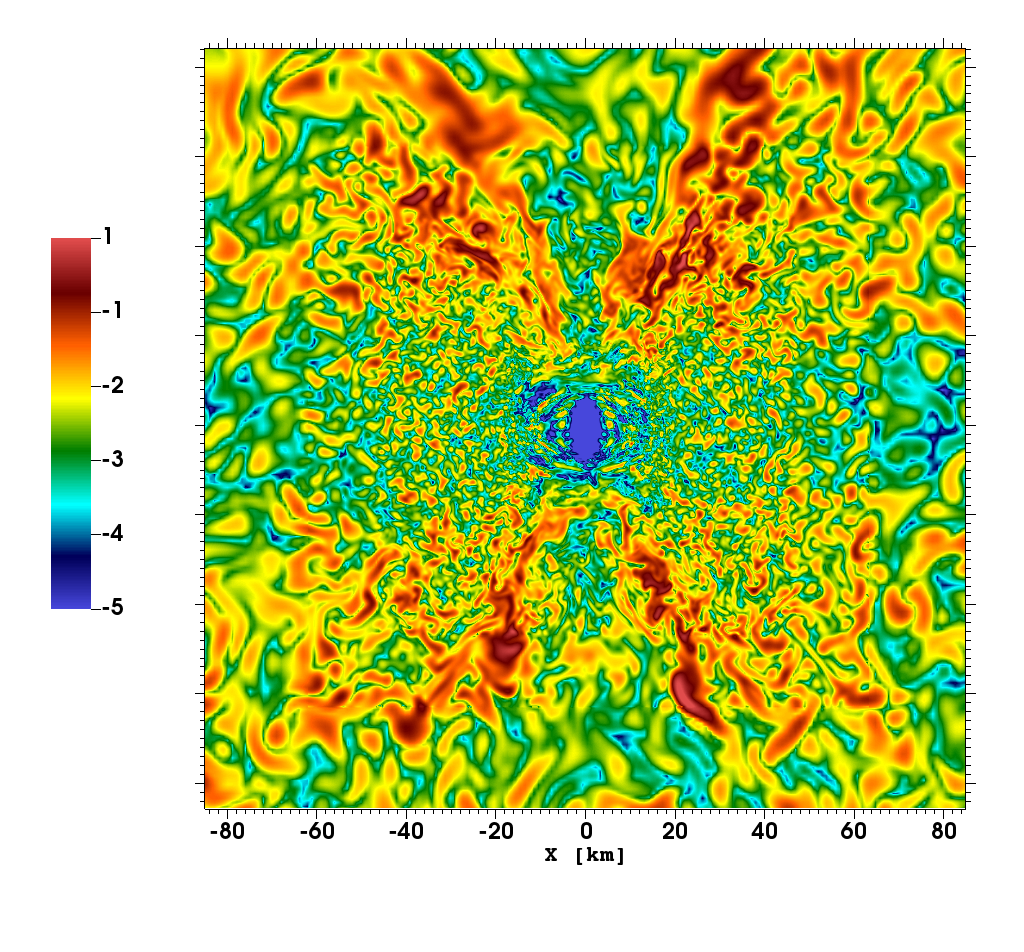}
	\caption{\textit{Comparison of different simulations at the meridional plane}. Snapshots at $50$ ms after the merger for the {\tt B15pol}, {\tt B15iso} and {\tt MRLES} (first, second and third column, respectively). Top: Magnetic field intensity in logarithmic scale. Bottom: Inverse of the plasma beta parameter. The {\tt B15iso} solution closely resembles the {\tt MRLES} one in distribution profile. In the {\tt B15pol} simulation, there is a significant excess of magnetic field near the axis, an artifact likely caused by the unrealistic large-scale initial field.}
	\label{fig:2D_plots_Bdiff}
\end{figure*}

The simulations performed to study the impact of the initial magnetic field are analyzed here. 
In the top panel of Fig.~\ref{fig:integrals}, the magnetic energies are displayed for the different initial conditions. Differences clearly arise among the different cases. The ${\tt B16pol}$ simulation reaches a value of $\sim10^{51}$~erg in few milliseconds, one order of magnitude larger than the ${\tt MRLES}$ case. After a small growing phase, it finally stabilizes at the saturation energy of $\sim 3 \times 10^{51}$~erg. On the other hand, the energy of ${\tt B14pol}$ remains approximately one order of magnitude lower with respect to the ${\tt MRLES}$ for all times. At the end of the exponentially growing phase, the magnetic energy of the remaining simulations (${\tt B15pol}$ and ${\tt B15iso}$) reaches values closer to the ones achieved in the ${\tt MRLES}$ case (i.e.,  between $\sim 10^{49}-10^{50}$~ergs). Afterwards, this energy grows mainly at the same rate for these three cases. This comparison between simulations not resolving the KHI with the ${\tt MRLES}$ case allow us to bracket a range of initial magnetic fields that could mimick the convergent results. Too high values, $\gtrsim 10^{15}$~G, will result in a too fast and strong amplification followed by prompt saturation of the magnetic energy, besides having a different configuration, that clearly remembers the unrealistic initial condition (as we show below). Low values $\lesssim 10^{14}$~G would lead to dynamics that would not match the energetics in under-resolved simulations, where the magnetic field does not grow significantly due to the KHI. Although these differences in the energy behaviour would already allow us to discard the cases ${\tt B16pol}$ and ${\tt B14pol}$, we will keep them in some of the analysis to stress the differences in the dynamics.

The volume-averaged magnetic fields computed within the bulk (middle panel) and in the envelope (bottom panel) are plotted as a function of time for the different cases. 
%The dashed and dotted lines correspond to the toroidal component and the poloidal one, respectively.
Focusing on the middle plot, all simulations are toroidally dominated from $\sim 10$ ms after the merger. As it was already shown in the top plot, the results from both ${\tt B16pol}$ and ${\tt B14pol}$ simulations differ significantly from the ${\tt MRLES}$ simulation. On the other hand, the toroidal magnetic field of ${\tt B15pol}$ and ${\tt B15iso}$ behaves in a comparable way to the one in ${\tt MRLES}$, and differs only by a factor of $2-3$ which can be easily adjusted by rescaling the initial data. The poloidal component of these three simulations decreases after the exponential amplification phase, while still maintaining the same shape. The bottom plot shows a similar behaviour between these three simulations (i.e., ${\tt B15pol}$, ${\tt B15iso}$ and${\tt MRLES}$) for the poloidal component, again within a factor of a few. However, the toroidal magnetic field in the envelope continues to decrease at $t\sim 50$~ms, a trend not replicated by any of our underesolved simulations. 
%\FC{hara falta aclarar aqui (e.g. en una footnote) que no habiamos demostrado convergencia en el envelope?}
%CP: que no hayamos encontrado saturacion no significa que los valores fueran completamente locos. Estaban dentro de un factor 2, asi que el trend si que creo que es fiable. Mejor no liar.

From the above discussion, it is evident that only the simulations {\tt B15pol} and {\tt B15iso} are close to our convergent {\tt MRLES} results. To visually compare these simulations, we plot the total magnetic field and the inverse of the plasma beta parameter at $50$~ms after the merger in Fig.~\ref{fig:2D_plots_Bdiff} (top and bottom rows, respectively) for the {\tt B15pol}, {\tt B15iso} and {\tt MRLES} (first, second and third columns, respectively). The {\tt B15iso} behaves similar to the {\tt MRLES} in both quantities. However, the {\tt B15pol} exhibits a more intense magnetic field above and below the remnant, strongly indicating the formation of magnetically dominated helicoidal structures associated with the jet. We want to stress that no jet has been observed in the {\tt MRLES} simulation at $250$~ms after the merger. Although the volume-averaged magnetic field of the {\tt B15pol} simulation has similar values with respect to our convergent {\tt MRLES}, its behaviour is completely different. We suggest that, if a jet is launched in this simulation, it would likely be due to the biased memory of the initial strong large-scale, purely poloidal,  magnetic field configuration.

\begin{figure*}[ht!]
\centering
	\includegraphics[width=\linewidth]{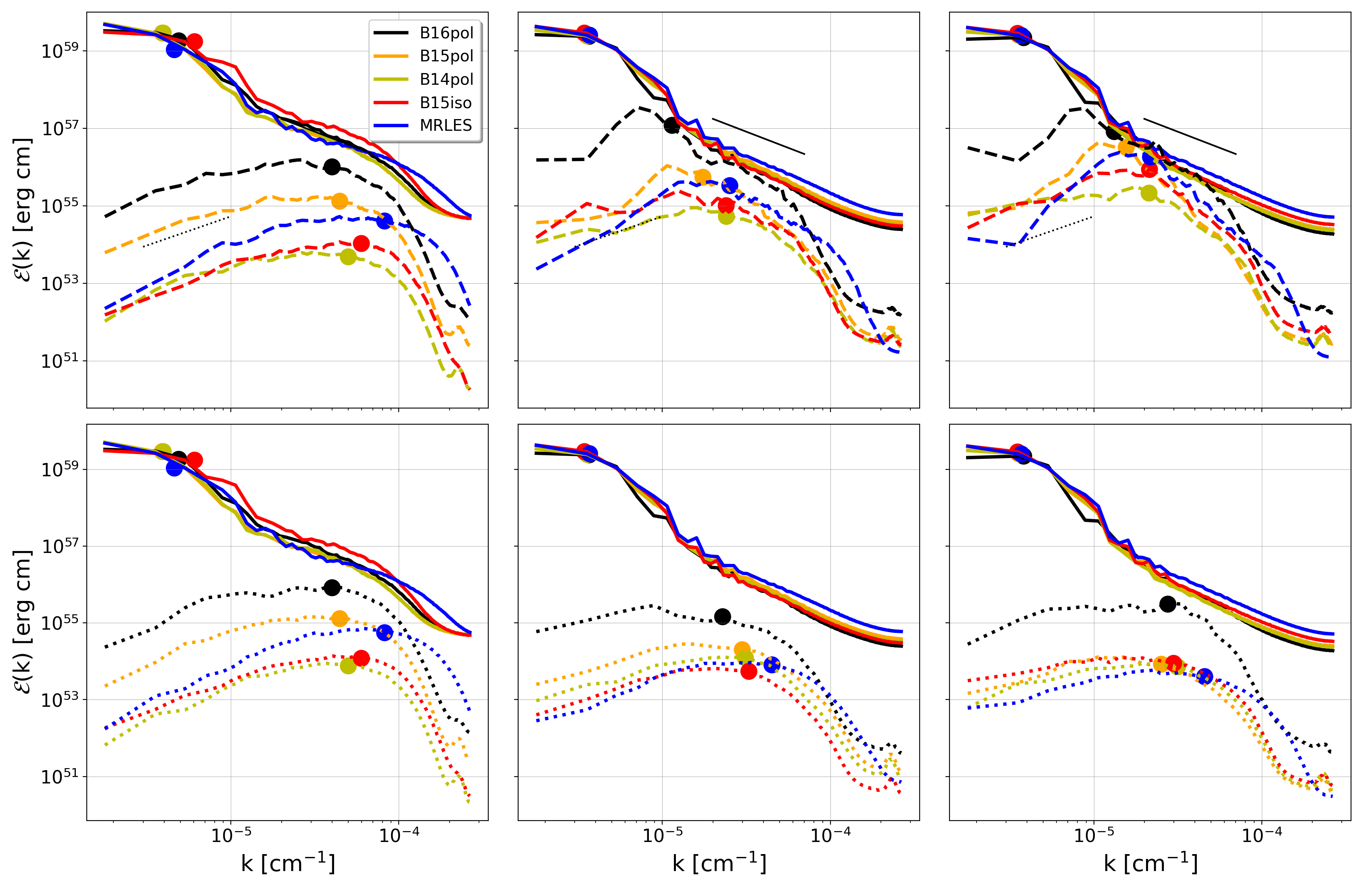}
	\caption{\textit{Spectral energy evolution}. Kinetic (solid) and toroidal (top) and poloidal (bottom) magnetic energy spectra (dashed and dotted, respectively) as a function of the wavenumber. Different colors represent different simulations. From left to right, $t = \{5,25,50\}$ ms after the merger. Dotted and solid straight lines correspond to Kazantsev ($k^{3/2}$) and Kolmogorov ($k^{-5/3}$) slopes, respectively. The magnetic spectra for the cases ${\tt B16pol}$ and ${\tt B14pol}$, does not match the ${\tt MRLES}$ results, especially in the toroidal components.}
	\label{fig:spectra_all}
\end{figure*}

\begin{figure}[h!]
\centering
\includegraphics[width=1\linewidth]{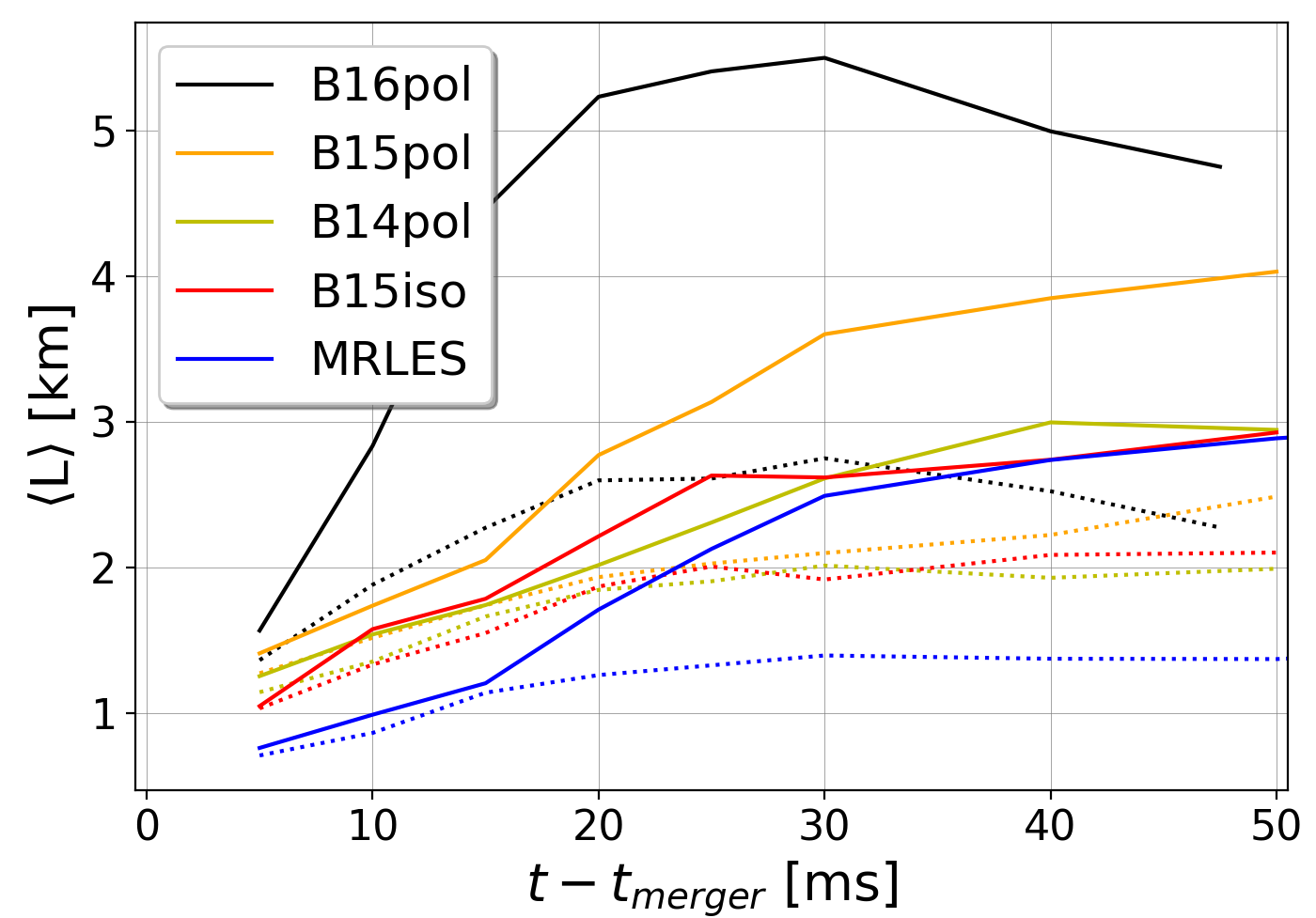}
\caption{\textit{Average characteristic lengthscale}. 
	Average characteristic length of toroidal (solid) and poloidal (dotted) components of the magnetic field as a function of time for the different simulations, defined as $\langle L \rangle = 2\pi/ \langle k \rangle $. The best match to the ${\tt MRLES}$ case is obtained with the ${\tt B15iso}$ simulation.}
\label{fig:Average_characteristic_wavelength}
\end{figure}

A more in-depth analysis is achieved by computing the spectral energies. In Fig~\ref{fig:spectra_all} we plot the spectra of all simulations at $t=\{5, 25, 50\}$~ms after the merger. Both rows show the kinetic energy, with the top row displaying the toroidal component of the magnetic energy and the bottom row displaying the poloidal part. Following the previously sketched pattern, the ${\tt B16pol}$ simulation largely differs from the other simulations from the very beginning, with a toroidal component quickly reaching equipartition with the kinetic energy at intermediate scales and the poloidal component almost reaching equipartition after $t = 50$~ms. As expected, both components of the magnetic energy in the ${\tt B15pol}$, ${\tt B15iso}$ and ${\tt MRLES}$ cases behave similarly from $25$~ms on. 

Interesting information from the spectra can be obtained by computing the average characteristic wavenumber, displayed  in Fig.~\ref{fig:Average_characteristic_wavelength}, indicates the spatial scale at which most of the energy concentrates. 
The solid lines correspond to the toroidal points of the first row of Fig~\ref{fig:spectra_all}, and the dotted lines correspond to the poloidal ones. We can now discern that the toroidal component of ${\tt B15pol}$ is significantly larger than the one of ${\tt MRLES}$. On the other hand, there is a very good match between the ${\tt B15iso}$ and the ${\tt MRLES}$ cases in the toroidal component from $\sim 30$~ms on. Although none of the cases perfectly matches the poloidal component of ${\tt MRLES}$, it is again the ${\tt B15iso}$ the simulation that comes closer.

These results strongly suggest that, in order to model effectively the turbulent amplification phase by setting an unrealistically strong initial magnetic field (i.e., such that the evolved magnetic field best matches our convergent ${\tt MRLES}$ solutions) when using a moderate resolution $\Delta x \approx 120$~m, the best option is to start with a strong, isotropic magnetic field like the one in the ${\tt B15iso}$ case.

%%%%%%%%%%%%%%%%%%%%%%%%%%%%%%%%%%%%%%%%%%%%%
\section{Conclusions} \label{sec:conclusions}
%%%%%%%%%%%%%%%%%%%%%%%%%%%%%%%%%%%%%%%%%%%%%

In this paper, we have studied the long-time dynamics of the HMNS remnant produced by the merger of two magnetized neutron stars. We have performed a high-resolution Large-Eddy-Simulation of the coalescence, where the individual neutron stars contain initial realistic magnetic fields $B \sim 10^{11}$~G. Within this setup, we continued a previous simulation from $100$~ms to a final time of $250$~ms after the merger. We have observed the development of large-scale, strong magnetic helicoidal structures, especially after $t = 150$~ms. However, we did not find any evidence of jet launching on the timescales spanned by our simulation. The fact that other works using a similar numerical setup (i.e., atmosphere density, numerical schemes,...) but much stronger initial magnetic fields (see for instance~\cite{ciolfi2020collimated,combi23,2024NatAs...8..298K,2024arXiv240503705B}) found jets on comparable timescales suggests that unrealistic choices  might facilitate the jet formation. One of the main conclusions of this work is that, when realistic magnetic fields are considered, it requires longer times to form  large coherent structures, thus delaying the jet formation. Unfortunately, at such long timescales neutrino cooling and re-heating is expected to play a crucial role, so it is pointless to continue our simulations beyond these timescales without including them.

The dependence of the initial magnetic field on the jet formation, in simulations with moderate resolutions which can not resolve the KHI, is studied by performing several simulations without LES changing both the strength and the topology of the initial magnetic field. Our results suggest that, with poloidal topologies and using high-order HRSC numerical schemes with resolution of $120$~m, the energetics is only recovered for initial strength range of $ 10^{14}~\textrm{G} \lesssim B \lesssim 10^{15}~\textrm{G}$. %\DV{\bf This range of value might be highly misleading... It strongly depend on the resolution and the scheme: either specify that they are for our case and res., or be more vague in the range.}.
Snapshots of the magnetic field strength on the meridional plane shows that, even within those bounds, the poloidal initial data leads to larger magnetization near the rotational axis than our resolved ${\tt MRLES}$ case. A closer analysis of the spectral distribution also reveals that the characteristic length-scales are still significantly larger than in the resolved ${\tt MRLES}$ case. An initial condition given by a strong but isotropic small-scale field, instead of a poloidal large-scale one, allow us to match reasonably well both the energetics and the spectra, including the characteristic scale, as already suggested in Ref.~\cite{aguilera23}.

%%%%%%%%%%%%%%%%%%%%%%%%%%%%%%%
\subsection*{Acknowledgements}
%%%%%%%%%%%%%%%%%%%%%%%%%%%%%%%

This work was supported by the Grant PID2022-138963NB-I00 funded by MCIN/AEI/10.13039/501100011033/FEDER, UE. RA-M is funded by he Deutsche Forschungsgemeinschaft (DFG, German Research Foundation) under Germany’s Excellence Strategy – EXC 2121 ”Quantum Universe” – 390833306. FC acknowledge financial support from CONICET, SeCyT-UNC, and MinCyT-Argentina. DV is funded by the European Research Council (ERC) Starting Grant IMAGINE (grant agreement No. [948582]) under the European Union’s Horizon 2020 research and innovation programme. DV's work was also partially supported by the program Unidad de Excelencia María de Maeztu CEX2020-001058-M. SR has been supported by the Swedish Research Council (VR) under grant number 2020-05044, by the research environment grant “Gravitational Radiation and Electromagnetic Astrophysical Transients” (GREAT) funded by the Swedish Research Council (VR) under Dnr 2016-06012, by the Knut and Alice Wallenberg Foundation under grant Dnr. KAW 2019.0112, by Deutsche Forschungsgemeinschaft (DFG, German Research Foundation) under Germany’s Excellence Strategy - EXC 2121 “Quantum Universe” - 390833306 and by the European Research Council (ERC) Advanced Grant INSPIRATION under the European Union’s Horizon 2020 research and innovation programme (Grant agreement No. 101053985). The authors thankfully acknowledges RES resources provided by BSC in MareNostrum to RES-AECT-2023-2-0002 and RES-AECT-2024-1-0010 and  the facilities of the North-German Supercomputing Alliance (HLRN).

%%%%%%%%%%%%%%%%%%%%%%%%%%%%%REFERENCES
\bibliographystyle{unsrt}
\bibliography{turbulence_new}

\begin{thebibliography}{10}

\bibitem{LVC-BNS}
B.P. Abbott et~al.
\newblock Gw170817: Observation of gravitational waves from a binary neutron
  star inspiral.
\newblock {\em Phys. Rev. Lett.}, 119:161101, 2017.

\bibitem{LVC-MMA}
B.P. Abbott et~al.
\newblock {Multi-messenger Observations of a Binary Neutron Star Merger}.
\newblock {\em Astrophys. J. Lett.}, 848:L12, 2017.

\bibitem{LVC-GRB}
B.P. Abbott et~al.
\newblock Gravitational waves and gamma-rays from a binary neutron star merger:
  Gw170817 and grb 170817a.
\newblock {\em Astrophys. J. Lett.}, 848(2):L13, 2017.

\bibitem{goldstein2017}
A.~Goldstein et~al.
\newblock An ordinary short gamma-ray burst with extraordinary implications:
  Fermi-gbm detection of grb 170817a.
\newblock {\em Astrophys. J. Lett.}, 848(2):L14, 2017.

\bibitem{savchenko2017}
V.~Savchenko et~al.
\newblock Integral detection of the first prompt gamma-ray signal coincident
  with the gravitational-wave event gw170817.
\newblock {\em Astrophys. J. Lett.}, 848(2):L15, 2017.

\bibitem{Troja2017}
E.~Troja et~al.
\newblock The x-ray counterpart to the gravitational-wave event gw170817.
\newblock {\em Nature}, 551(7678):71--74, 2017.

\bibitem{Margutti2017}
R.~Margutti et~al.
\newblock The electromagnetic counterpart of the binary neutron star merger
  ligo/virgo gw170817. v. rising x-ray emission from an off-axis jet.
\newblock {\em Astrophys. J. Lett.}, 848(2):L20, 2017.

\bibitem{Hallinan2017}
G.~Hallinan et~al.
\newblock A radio counterpart to a neutron star merger.
\newblock {\em Science}, 358(6370):1579--1583, 2017.

\bibitem{Alexander2017}
K.~D. Alexander et~al.
\newblock The electromagnetic counterpart of the binary neutron star merger
  ligo/virgo gw170817. vi. radio constraints on a relativistic jet and
  predictions for late-time emission from the kilonova ejecta.
\newblock {\em Astrophys. J. Lett.}, 848(2):L21, 2017.

\bibitem{Mooley2018a}
K.P. Mooley et~al.
\newblock A mildly relativistic wide-angle outflow in the neutron-star merger
  event gw170817.
\newblock {\em Nature}, 554(7691):207--210, 2018.

\bibitem{Lazzati2018}
D.~Lazzati, R.~Perna, B.J. Morsony, D.~Lopez-Camara, M.~Cantiello, R.~Ciolfi,
  B.~Giacomazzo, and J.C. Workman.
\newblock Late time afterglow observations reveal a collimated relativistic jet
  in the ejecta of the binary neutron star merger gw170817.
\newblock {\em Phys. Rev. Lett.}, 120:241103, 2018.

\bibitem{Lyman2018}
J.D. Lyman et~al.
\newblock The optical afterglow of the short gamma-ray burst associated with
  gw170817.
\newblock {\em Nature Astr.}, 2(9):751--754, 2018.

\bibitem{Alexander2018}
K.D. Alexander et~al.
\newblock A decline in the x-ray through radio emission from gw170817 continues
  to support an off-axis structured jet.
\newblock {\em Astrophys. J. Lett.}, 863(2):L18, 2018.

\bibitem{Mooley2018b}
K.P. Mooley et~al.
\newblock Superluminal motion of a relativistic jet in the neutron-star merger
  gw170817.
\newblock {\em Nature}, 561(7723):355--359, 2018.

\bibitem{Ghirlanda2019}
G.~Ghirlanda et~al.
\newblock Compact radio emission indicates a structured jet was produced by a
  binary neutron star merger.
\newblock {\em Science}, 363(6430):968--971, 2019.

\bibitem{lattimer74}
J.M. Lattimer and D.N. Schramm.
\newblock {Black-Hole-Neutron-Star Collisions}.
\newblock {\em Astrophys. J. Lett.}, 192:L145, 1974.

\bibitem{eichler89}
D.~Eichler, M.~Livio, T.~Piran, and D.N. Schramm.
\newblock {Nucleosynthesis, neutrino bursts and {\ensuremath{\gamma}}-rays from
  coalescing neutron stars}.
\newblock {\em Nature}, 340(6229):126--128, 1989.

\bibitem{rosswog99}
S.~Rosswog, M.~Liebend{\"o}rfer, F.~K. Thielemann, M.B. Davies, W.~Benz, and
  T.~Piran.
\newblock {Mass ejection in neutron star mergers}.
\newblock {\em J. Astrophys. Astron.}, 341:499--526, 1999.

\bibitem{freiburghaus99}
C.~Freiburghaus, S.~Rosswog, and F.K. Thielemann.
\newblock r-process in neutron star mergers.
\newblock {\em Astrophys. J.}, 525(2):L121, 1999.

\bibitem{Arcavi2017}
I.~Arcavi et~al.
\newblock Optical emission from a kilonova following a
  gravitational-wave-detected neutron-star merger.
\newblock {\em Nature}, 551(7678):64--66, 2017.

\bibitem{Coulter2017}
D.A. Coulter et~al.
\newblock Swope supernova survey 2017a (sss17a), the optical counterpart to a
  gravitational wave source.
\newblock {\em Science}, 358(6370):1556--1558, 2017.

\bibitem{Pian2017}
E.~Pian et~al.
\newblock Spectroscopic identification of r-process nucleosynthesis in a double
  neutron-star merger.
\newblock {\em Nature}, 551(7678):67--70, 2017.

\bibitem{Smartt2017}
S.J. Smartt et~al.
\newblock A kilonova as the electromagnetic counterpart to a gravitational-wave
  source.
\newblock {\em Nature}, 551(7678):75--79, 2017.

\bibitem{Kasen2017}
D.~Kasen, B.~Metzger, J.~Barnes, E.~Quataert, and E.~Ramirez-Ruiz.
\newblock Origin of the heavy elements in binary neutron-star mergers from a
  gravitational-wave event.
\newblock {\em Nature}, 551(7678):80--84, 2017.

\bibitem{Metzger2019LRR}
B.D. Metzger.
\newblock Kilonovae.
\newblock {\em Liv. Rev. Rel.}, 23(1):1, 2020.

\bibitem{kiuchi15}
K.~Kiuchi, P.~Cerd\'a-Dur\'an, K.~Kyutoku, Y.~Sekiguchi, and M.~Shibata.
\newblock Efficient magnetic-field amplification due to the kelvin-helmholtz
  instability in binary neutron star mergers.
\newblock {\em Phys. Rev. D}, 92:124034, 2015.

\bibitem{palenzuela22}
C.~Palenzuela, R.~Aguilera-Miret, F.~Carrasco, R.~Ciolfi, J.~V. Kalinani,
  W.~Kastaun, B.~Mi\~nano, and D.~Vigan\`o.
\newblock Turbulent magnetic field amplification in binary neutron star
  mergers.
\newblock {\em Phys. Rev. D}, 106:023013, 2022.

\bibitem{2024NatAs...8..298K}
K.~Kiuchi, A.~Reboul-Salze, M.~Shibata, and Y.~Sekiguchi.
\newblock A large-scale magnetic field produced by a solar-like dynamo in
  binary neutron star mergers.
\newblock {\em Nature Astr.}, 8(3):298--307, 2024.

\bibitem{aguilera23}
R.~Aguilera-Miret, C.~Palenzuela, F.~Carrasco, and D.~Vigan\`o.
\newblock Role of turbulence and winding in the development of large-scale,
  strong magnetic fields in long-lived remnants of binary neutron star mergers.
\newblock {\em Phys. Rev. D}, 108:103001, 2023.

\bibitem{ciolfi2020key}
R.~Ciolfi.
\newblock The key role of magnetic fields in binary neutron star mergers.
\newblock {\em Gen. Relativ. Gravit.}, 52(6):59, 2020.

\bibitem{Kiuchireview}
K.~Kiuchi.
\newblock General relativistic magnetohydrodynamics simulations for binary
  neutron star mergers.
\newblock 2024.

\bibitem{ruiz16}
M.~Ruiz, R.N. Lang, V.~Paschalidis, and S.L. Shapiro.
\newblock Binary neutron star mergers: A jet engine for short gamma-ray bursts.
\newblock {\em Astrophys. J. Lett.}, 824(1):L6, 2016.

\bibitem{ciolfi2020collimated}
R.~Ciolfi.
\newblock Collimated outflows from long-lived binary neutron star merger
  remnants.
\newblock {\em MNRAS: Letters}, 495(1):L66--L70, 2020.

\bibitem{aguilera22}
R.~Aguilera-Miret, D.~Viganò, and C.~Palenzuela.
\newblock Universality of the turbulent magnetic field in hypermassive neutron
  stars produced by binary mergers.
\newblock {\em Astrophys. J. Lett.}, 926(2):L31, 2022.

\bibitem{alic12}
D.~Alic, C.~Bona-Casas, C.~Bona, L.~Rezzolla, and C.~Palenzuela.
\newblock Conformal and covariant formulation of the z4 system with
  constraint-violation damping.
\newblock {\em Phys. Rev. D}, 85:064040, 2012.

\bibitem{bezares17}
M.~Bezares, C.~Palenzuela, and C.~Bona.
\newblock Final fate of compact boson star mergers.
\newblock {\em Phys. Rev. D}, 95:124005, 2017.

\bibitem{shibatabook}
M.~Shibata.
\newblock {\em {Numerical Relativity}}.
\newblock 2016.

\bibitem{palenzuela15}
C.~Palenzuela, S.L. Liebling, D.~Neilsen, L.~Lehner, O.L. Caballero,
  E.~O'Connor, and M.~Anderson.
\newblock Effects of the microphysical equation of state in the mergers of
  magnetized neutron stars with neutrino cooling.
\newblock {\em Phys. Rev. D}, 92:044045, 2015.

\bibitem{read09}
J.S. Read, B.D. Lackey, B.J. Owen, and J.L. Friedman.
\newblock Constraints on a phenomenologically parametrized neutron-star
  equation of state.
\newblock {\em Phys. Rev. D}, 79:124032, 2009.

\bibitem{Endrizzi2016}
A.~Endrizzi, R.~Ciolfi, B.~Giacomazzo, W.~Kastaun, and T.~Kawamura.
\newblock General relativistic magnetohydrodynamic simulations of binary
  neutron star mergers with the apr4 equation of state.
\newblock {\em Class. Quantum Grav.}, 33(16):164001, 2016.

\bibitem{kastaun20}
W.~Kastaun, J.V. Kalinani, and R.~Ciolfi.
\newblock Robust recovery of primitive variables in relativistic ideal
  magnetohydrodynamics.
\newblock {\em Phys. Rev. D}, 103:023018, 2021.

\bibitem{vigano20}
D.~Vigan\`o, R.~Aguilera-Miret, F.~Carrasco, B.~Mi\~nano, and C.~Palenzuela.
\newblock General relativistic mhd large eddy simulations with gradient
  subgrid-scale model.
\newblock {\em Phys. Rev. D}, 101:123019, 2020.

\bibitem{aguilera2020}
R.~Aguilera-Miret, D.~Vigan\`o, F.~Carrasco, B.~Mi\~nano, and C.~Palenzuela.
\newblock Turbulent magnetic-field amplification in the first 10 milliseconds
  after a binary neutron star merger: Comparing high-resolution and large-eddy
  simulations.
\newblock {\em Phys. Rev. D}, 102:103006, 2020.

\bibitem{arbona13}
A.~Arbona, A.~Artigues, C.~Bona-Casas, J.~Massó, B.~Miñano, A.~Rigo,
  M.~Trias, and C.~Bona.
\newblock Simflowny: A general-purpose platform for the management of physical
  models and simulation problems.
\newblock {\em Comput. Phys. Commun.}, 184(10):2321--2331, 2013.

\bibitem{arbona18}
A.~Arbona, B.~Miñano, A.~Rigo, C.~Bona, C.~Palenzuela, A.~Artigues,
  C.~Bona-Casas, and J.~Massó.
\newblock Simflowny 2: An upgraded platform for scientific modelling and
  simulation.
\newblock {\em Comput. Phys. Commun.}, 229:170--181, 2018.

\bibitem{hornung02}
R.D. Hornung and S.R. Kohn.
\newblock Managing application complexity in the samrai object-oriented
  framework.
\newblock {\em Concurr. Comp. Pract. E.}, 14(5):347--368, 2002.

\bibitem{gunney16}
Brian T.~N. Gunney and R.~W. Anderson.
\newblock Advances in patch-based adaptive mesh refinement scalability.
\newblock {\em J. Parallel. Distr. Com.}, 89:65 -- 84, 2016.

\bibitem{shu98}
Chi-Wang Shu.
\newblock {\em Essentially non-oscillatory and weighted essentially
  non-oscillatory schemes for hyperbolic conservation laws}, pages 325--432.
\newblock Springer Berlin Heidelberg, 1998.

\bibitem{suresh97}
A.~Suresh and H.T. Huynh.
\newblock Accurate monotonicity-preserving schemes with runge–kutta time
  stepping.
\newblock {\em J. Comput. Phys.}, 136(1):83 -- 99, 1997.

\bibitem{McCorquodale:2011}
P.~McCorquodale and P.~Colella.
\newblock A high-order finite-volume method for conservation laws on locally
  refined grids.
\newblock {\em Commun. Appl. Math. Comput. Sci.}, 6(1):1--25, 2011.

\bibitem{Mongwane:2015}
B.~Mongwane.
\newblock Toward a consistent framework for high order mesh refinement schemes
  in numerical relativity.
\newblock {\em Gen. Relativ. Gravit.}, 47:1--21, 2015.

\bibitem{palenzuela18}
C.~Palenzuela, B.~Miñano, D.~Viganò, A.~Arbona, C.~Bona-Casas, A.~Rigo,
  M.~Bezares, C.~Bona, and J.~Massó.
\newblock A simflowny-based finite-difference code for high-performance
  computing in numerical relativity.
\newblock {\em Class. Quantum Grav.}, 35(18):185007, 2018.

\bibitem{vigano19}
D.~Viganò, D.~Martínez-Gómez, J.A. Pons, C.~Palenzuela, F.~Carrasco,
  B.~Miñano, A.~Arbona, C.~Bona, and J.~Massó.
\newblock A simflowny-based high-performance 3d code for the generalized
  induction equation.
\newblock {\em Comput. Phys. Commun.}, 237:168--183, 2019.

\bibitem{lorene}
{\sc Lorene}.
\newblock \url{http://www.lorene.obspm.fr/}, 2010.

\bibitem{LVC-170817properties}
B.P. Abbott et~al.
\newblock Properties of the binary neutron star merger gw170817.
\newblock {\em Phys. Rev. X}, 9:011001, 2019.

\bibitem{kiuchi18}
K.~Kiuchi, K.~Kyutoku, Y.~Sekiguchi, and M.~Shibata.
\newblock Global simulations of strongly magnetized remnant massive neutron
  stars formed in binary neutron star mergers.
\newblock {\em Phys. Rev. D}, 97:124039, 2018.

\bibitem{ciolfi2019}
R.~Ciolfi, W.~Kastaun, J.V. Kalinani, and B.~Giacomazzo.
\newblock First 100 ms of a long-lived magnetized neutron star formed in a
  binary neutron star merger.
\newblock {\em Phys. Rev. D}, 100:023005, 2019.

\bibitem{ruiz2020}
M.~Ruiz, A.~Tsokaros, and S.L. Shapiro.
\newblock Magnetohydrodynamic simulations of binary neutron star mergers in
  general relativity: Effects of magnetic field orientation on jet launching.
\newblock {\em Phys. Rev. D}, 101:064042, 2020.

\bibitem{bahramian23}
A.~Bahramian and N.~Degenaar.
\newblock {\em Low-Mass X-ray Binaries}, pages 1--62.
\newblock Springer Nature Singapore, 2022.

\bibitem{balbus91}
S.A. Balbus and J.F. Hawley.
\newblock A powerful local shear instability in weakly magnetized disks. i -
  linear analysis. ii - nonlinear evolution.
\newblock {\em Astrophys. J.}, 376:214--233, 1991.

\bibitem{combi23}
L.~Combi and D.M. Siegel.
\newblock Jets from neutron-star merger remnants and massive blue kilonovae.
\newblock {\em Phys. Rev. Lett.}, 131:231402, 2023.

\bibitem{2024arXiv240503705B}
J.~Bamber, A.~Tsokaros, M.~Ruiz, and S.L. Shapiro.
\newblock Jetlike structures in low-mass binary neutron star merger remnants.
\newblock {\em Phys. Rev. D}, 110:024046, 2024.

\end{thebibliography}
%%%%%%%%%%%%%%%%%%%%%%%%%%%%%

\end{document}